\documentclass{emulateapj}

\newcommand{\ebv}{$E(B-V)$}

\newcommand{\Msun}{ $M_{\odot}$}

\newcommand{\zs}{$z$$\sim$}
\newcommand{\sqam}{arcmin$^2$}
\newcommand{\Ebv}{$E(B-V)$}
\newcommand{\OmOlHo}{$(\Omega_M, \Omega_\Lambda, H_0)$}
\newcommand{\cosmoparams}{0.3, 0.7, 70 km s$^{-1}$ Mpc$^{-1}$}

\newcommand{\UGRI}{$U_n G {\cal R} I$}

\newcommand{\G}{$G$}
\newcommand{\R}{$\cal R$}
\newcommand{\I}{$I$}
\newcommand{\Rlim}{${\cal R}_{lim}$}
\newcommand{\GRave}{$G{\cal R}$}
\newcommand{\UG}{$U_n-G$}

\newcommand{\GR}{$G-{\cal R}$}
\newcommand{\RI}{${\cal R}-I$}
\newcommand{\UGR}{$U_n G {\cal R}$}
\newcommand{\GRI}{$G {\cal R} I$}

\newcommand{\Mstar}{$M^*$}
\newcommand{\Lstar}{$L^*$}
\newcommand{\subLstar}{sub-$L^*$}
\newcommand{\phistar}{$\phi^*$}
\newcommand{\chisq}{$\chi^2$}
\newcommand{\Muv}{$M_{1700}$}
\newcommand{\Veff}{$V_{eff}$}

\newcommand{\phim}{$\phi(m)$}
\newcommand{\phiM}{$\phi(M)$}
\newcommand{\PhiM}{$\Phi(M)$}
\newcommand{\Phibar}{$\bar{\Phi}$}
\newcommand{\rootN}{$N^{1/2}$}
\newcommand{\pmz}{$p(m,z)$}

\newcommand{\kdfdata}{KDF~I}

\newcommand{\kdfsfhist}{KDF~III}
\newcommand{\kdfclustering}{KDF~IV}

\shorttitle{Keck Deep Fields. II. The UV Luminosity Function}
\shortauthors{Sawicki \& Thompson}

\begin{document}


\title{Keck Deep Fields. II. The Ultraviolet Galaxy Luminosity Function at 
\zs4, 3, and 2
\altaffilmark{1}}

\author{Marcin Sawicki\altaffilmark{2}
} 
\affil{
Dominion Astrophysical Observatory, Herzberg Institute of
Astrophysics, National Research Council, 5071~West Saanich Road,
Victoria, B.C., V9E 2E7, Canada; and Caltech Optical Observatories,
California Institute of Technology, MS 320-47, Pasadena, CA 91125, USA
}

\email{marcin.sawicki@nrc.gc.ca}

\author{David Thompson} 
\affil{
Caltech Optical Observatories, 
California Institute of Technology,
MS 320-47, Pasadena, CA 91125, 
USA}
\email{djt@irastro.caltech.edu}

\slugcomment{Accepted for publication in ApJ}

\altaffiltext{1}{Based on data obtained at the 
W.M.\ Keck Observatory, which is operated as a scientific partnership
among the California Institute of Technology, the University of
California, and NASA, and was made possible by the generous financial
support of the W.M.\ Keck Foundation.}
\altaffiltext{2}{Present address: Department of Physics, 
University of California, Santa Barbara, CA 93106, USA}

\begin{abstract}
We use very deep \UGRI\ multi-field imaging obtained at the Keck
telescope to study the evolution of the rest-frame 1700\AA\ galaxy
luminosity function as the Universe doubles its age from \zs4 to \zs2.
We use {\it exactly} the same filters and color-color selection as
those used by the Steidel team, but probe significantly fainter limits
--- well below \Lstar.  The depth of our imaging allows us to
constrain the faint end of the luminosity function reaching
\Muv$\sim$$-$18.5 at \zs3 (equivalent to $\sim$1\Msun/yr) accounting
for both \rootN\ uncertainty in the number of galaxies and for cosmic
variance.  We carefully examine many potential sources of systematic
bias in our LF measurements before drawing the following conclusions.
We find that the luminosity function of Lyman Break Galaxies evolves
with time and that this evolution is differential with luminosity.
The result is best constrained between the epochs at \zs4 and \zs3,
where we find that the number density of \subLstar\ galaxies increases
with time by at least a factor of 2.3 (11$\sigma$ statistical
confidence); while the faint end of the LF evolves, the bright end
appears to remain virtually unchanged, indicating that there may be
differential, luminosity-dependent evolution (98.5\% statistical
probability).  Potential systematic biases restrict our ability to
draw strong conclusions about continued evolution of the luminosity
function to lower redshifts, \zs2.2 and \zs1.7, but, nevertheless, it
appears certain that the number density of \zs2.2 galaxies at all
luminosities we studied, $-22$$>$\Muv$>$$-18$, is at least as high as
that of their counterparts at \zs3.  While it is not yet clear what
mechanism underlies the observed evolution, the fact that this
evolution is differential with luminosity opens up new avenues of
improving our understanding of how galaxies form and evolve at high
redshift.
\end{abstract}


\section{INTRODUCTION}

Understanding the formation and evolution of galaxies continues to be
one of the most active fields of observational cosmology.  Over the
last decade, advances in instrumentation and technique have made it
possible to select large samples of ``normal'' star-forming galaxies
for direct study at redshifts that correspond to a time when the
Universe was only a tenth of its present age and so study galaxy
assembly at a time when galaxies were young.  Several different
approaches for selecting high-$z$ galaxies are used, including
selection in rest-frame far-IR (e.g., Barger et al.\ 1998; Hughes et
al 1998; Blain et al.\ 1999; Eales et al.\ 2000), near-IR (e.g.,
Sawicki 2002), optical (e.g., Thompson et al.\ 1999; Cimatti et al.\
2002; Sawicki et al.\ 2005a) and the UV (e.g., Steidel et al.\ 1996,
1999, 2003, 2004; Sawicki, Lin, \& Yee 1997; Lowenthal et al.\ 1997;
Giavalisco 2002; Lehnert \& Bremer 2003; Stanway, Bunker \& McMahon
2003; Iwata et al. 2003; Ouchi et al. 2004a).  Among these different
techniques the Lyman Break Galaxy (LBG; Steidel et al.\ 1996, 2003)
surveys have yielded the largest, spectroscopically confirmed samples
and the most active and detailed follow-up studies.

Follow-up observations of LBG samples has taught us much about the
nature of these high-$z$ galaxies, although --- understandably ---
such follow-up has so far mainly focused on relatively luminous
objects at \zs3, where the samples are largest and spectroscopy is
easiest.  We now know, for example, that LBGs are dominated by fairly
young episodes of star formation and that they are enshrouded by large
amounts of interstellar dust (e.g., Sawicki \& Yee 1998; Ouchi et al.\
1999; Shapley et al.\ 2001; Papovich et al.\ 2001; Vijh, Witt, \&
Gordon 2003); that they are associated with massive dark matter halos
(Adelberger et al.\ 1998; Giavalisco et al.\ 1998); that they have
strong, starburst-driven outflows of material into the surrounding IGM
(Pettini et al.\ 1998, 2001, 2002; Adelberger et al. 2003); and that
they likely have sub-solar, but not primordial, metallicities (Pettini
et al.\ 2001, 2002).  However, most LBG studies have focused primarily
on relatively luminous objects ($L$$\gtrsim$\Lstar) at a single epoch
(\zs3), and comparisons of LBG properties as a function of time and
luminosity are still in their infancy.

Studying the properties of galaxies as a function of redshift has a
straightforward motivation rooted in the fact that observing galaxies
at different epochs allows us to study {\it directly} their evolution
as a function of time.  The initial search for and study of high-$z$
galaxies has been motivated by the desire to find the progenitors of
present-day galaxies and recent comparisons of galaxy populations {\it
between} different epochs beyond $z$$>$1 is its direct and natural
extension (e.g., Steidel et al. 1999; Adelberger et al.\ 2004; Ando et
al.\ 2004; Ferguson et al.\ 2004; Papovich et al.\ 2004).  Exploring
the time domain holds obvious but important attractions.

The study of high-$z$ galaxies as a function of their {\it luminosity}
is less obvious to motivate, especially given the fact that high-$z$
studies are observationally expensive even for luminous galaxies and
far more so for galaxies that are intrinsically faint.  Nevertheless,
a wealth of potential information lies to be discovered in comparisons
of galaxies as a function of luminosity.  The galaxy luminosity
function is not a simple power law as could be expected from the mass
function of dark matter halos (e.g., Jenkins et al.\ 2001), but
instead reflects the imprint of real differences in galaxy formation
and evolution processes.  If the luminosities of high-$z$ galaxies
correlate with the masses of their host dark matter halos --- as is
suggested by some clustering studies (Giavalisco \& Dickinson 2001;
Ouchi et al.\ 2004b) --- then the shape of the LF at high redshift
likely bears the direct imprint of star-formation driven feedback as a
function of halo mass. Additionally, luminosity may also reflect the
effects of processes such as fluctuating star formation rates ---
whether induced by galaxy-galaxy interactions or by other mechanisms
--- or differences in the properties of interstellar dust.  In any
case, it is unlikely that galaxies of different luminosity are just
trivially scaled copies of each other and so to fully understand the
story of galaxy formation we must study not just the brightest,
observationally most accessible members of the population, but also
their fainter cousins.  Such studies are particluarly attractive
because {\it differences} in evolution between galaxies with different
UV luminosities --- or, by extension, star formation rates --- can be
expected to point us to some of the most relevant mechanisms
responsible for driving galaxy evolution.  Finally, for any reasonable
luminosity function, most galaxies are sub-\Lstar galaxies and most of
the {\it luminosity} in the Universe is contained in galaxies below
\Lstar.  By extension, so is most of the star formation and
metal production activity.  The hitherto neglected sub-\Lstar\
high-redshift galaxies deserve our avid attention for all these
important reasons.

One of the most basic descriptors of a galaxy population is its
luminosity function (LF).  The shape of the galaxy luminosity function
bears the imprint of galaxy formation and evolution processes.  The
characteristic break in the LF seen at both low and high redshift
suggests that galaxies below \Lstar\ are not simple scaled replicas of
those above \Lstar\ but differ from them in more substantial ways.  At
present relatively little is known about the shape of the faint-end of
the LF of galaxies at high redshift, and about the evolution of the
high-$z$ LF.  What studies have been done are limited by small, often
single fields --- such as the HDF --- that can be affected by both
sample and cosmic variance (e.g., Sawicki, Lin, \& Yee 1997; Steidel
et al. 1999) or use samples whose fidelity has not been well tested
with spectroscopy (e.g., Iwata et al.\ 2003; Ouchi et al.\ 2004a;
Gabasch et al.\ 2004).

In this paper we use our large, very faint Keck Deep Fields (KDF)
galaxy samples to construct the luminosity functions of high-$z$
star-forming galaxies over a wide span of cosmic time (0.6, 0.8, and
0.8 Gyr from \zs4 to \zs3 to \zs2.2 to \zs1.7, respectively) and
reaching to very faint limits (\R=27, or
\Muv$\sim$$-$18, equivalent to star formation rate of
$\sim$1\Msun/yr).  At \zs4 and 3, we combine our samples with LF
measurements by Steidel et al.\ (1999) made from shallower, but
larger-area surveys to study the rest-frame UV-selected galaxy
luminosity function over up to a factor of a hundred in luminosity.

This paper is structured as follows.  In \S~\ref{introKDF} we briefly
describe our KDF sample of faint, UV-selected galaxies.  In
\S~\ref{LFcalculation} we describe the details of how we calculate the
LF and in \S~\ref{theLF} we describe our results and focus on
examining the possible sources of systematic error that may affect
them.  In \S~\ref{LFevolution} we examine several intriguing
evolutionary trends in the LF.  In \S~\ref{discussion} we discuss some
possible interpretations of the observed evolution and also point out
the potential new approaches to the study of galaxy evolution at high
redshift that the evolving LF opens to us.  Finally, in
\S~\ref{summary} we summarize our results.  As in all the papers in
the KDF series, we use the AB flux normalization (Oke, 1974) and adopt
$\Omega_M$=0.3, $\Omega_{\Lambda}$=0.7, and
$H_0$=70~km~s$^{-1}$~Mpc$^{-1}$.

\section{The Data}\label{introKDF}

Our study of the faint end of the high-redshift luminosity function
uses data from our very deep \UGRI\ Keck imaging survey --- the Keck
Deep Fields (KDFs).  The KDF was specifically designed to explore the
evolution of the population of {\it very} faint (sub-\Lstar)
star-forming galaxies at high redshift, including their luminosity
function and luminosity-dependent clustering.  The KDF probes the
hitherto poorly-explored faint end of the galaxy population at
redshifts \zs4 -- 1.7 by extending to fainter magnitudes the
well-known color selection techniques used by Steidel et al.\ (1999,
2003, 2004).  The KDF survey is described in detail in the companion
paper by Sawicki \& Thompson (2005; hereafter \kdfdata), which gives a
detailed description of the observations, data reduction, and
selection of star-forming galaxy samples at \zs2 -- 4.  Here, we give
only a brief overview of its main characteristics.

The KDF survey uses the {\it very same } \UGRI\ filter set and
color-color selection criteria that are used so successfully by
Steidel et al.\ (1999, 2003, 2004) to select their high-redshift
samples, but reaches a limiting magnitude \Rlim=27, that is 1.5
magnitudes deeper than the surveys by the Steidel group.  Because of
our use of an identical \UGRI\ filter set, our KDF data are a direct
and straightforward extension to fainter magnitudes of the
spectroscopically-tested and well-understood samples of Steidel et al.
Because of the extensive spectroscopic work of the Steidel team,
selection effects, including foreground interloper fractions (which
are only a few percent), are well known and can be safely used for our
fainter samples as explained in \kdfdata.

The KDF cover a total area of 169 \sqam\ and consist of five fields
(called fields 02A, 03A, 03B, 09A, and 09B) that were observed
separately and so reach slightly different depths ranging over
\Rlim$\sim$26.7--27.3 (50\% completeness).  The five fields are 
grouped into three patches (patches 02, 03, and 09) that are spatially
well-separated on the sky.  The division of the KDF into these three
spatially-independent patches gives us the extremely important ability
to monitor the magnitude and impact of cosmic variance.

Our ground-based images have typical seeing of $\sim$1\arcsec\,
ensuring that high-$z$ galaxies are unresolved and can be treated as
point sources.  This relatively poor spatial resolution is a blessing
in disguise as it drastically reduces concerns about galaxy size
selection biases that are present in observations with better image
quality, such as HST data.

Photometry is done in a manner virtually identical to that of the
Steidel et al.\ work, namely with object detection in very deep
\R-band images and color measurement through matched 2\arcsec-diameter
apertures on images smoothed to a common seeing.  Our \UGRI\ filter
set --- the same as that used by the Steidel team --- allows us to
select high-$z$ galaxies in a manner that is {\it identical} to their
brighter samples (see \kdfdata\ or Steidel et al.\ 1999, 2003, 2004,
for details of the selection criteria).  The photometric completeness
of our survey --- tested carefully using simulations --- is similar at
\R=27 to that of the Steidel et al.\ surveys at
\R=25.5.  The KDF therefore probes a factor of 4 deeper in luminosity
than the work of the Steidel team.  

To its nominal completeness limit of \Rlim=27, the KDF contain 427
\GRI-selected \zs4 Lyman Break Galaxies (LBGs), 1481
\UGR-selected \zs3 LBGs, 2417 \UGR-selected \zs2.2 star-forming
galaxies, and 2043 \UGR-selected \zs1.7 star-forming galaxies.

\section{CALCULATION OF THE LUMINOSITY FUNCTION}\label{LFcalculation}

We use the effective volume, \Veff, approach to compute the luminosity
functions.  Our approach is virtually identical to that used by
Steidel et al.\ (1999) on their brighter \zs3 and \zs4 samples.  This
section of the paper is devoted to a detailed discussion of the
technique we use to calculate the LF.  We defer the discussion of our
actual LF results to
\S~\ref{theLF} and later. 

The \Veff\ approach to calculating the LF is straightforward.  In
brief, (1) for each redshift sample (\zs4, \zs3, \zs2.2, and \zs1.7)
we first use simulations to determine the effective volumes, \Veff, of
the survey as a function of apparent magnitude --- volumes that
account for incompleteness due to objects missing from the sample.  We
then (2) combine these \Veff\ with the observed galaxy counts to
compute the incompleteness-corrected number density of galaxies at
each redshift as a function of apparent magnitude. And finally, (3) we
convert these apparent-magnitude number densities into
absolute-magnitude ones to arrive at the LFs.  These three steps are
described in detail in \S\S~\ref{Veffcalc}, \ref{numbercounts}, and
\ref{absmags}, respectively

We measure the LF at rest-frame 1700\AA\ for two reasons.  Because (1)
LF calculation at this rest-frame wavelength matches the LF analysis
by Steidel et al.\ (1999) using brighter \zs3 and \zs4 Lyman Break
Galaxy samples.  But also because (2) --- as we will see in more
detail in \S~\ref{absmags} --- rest-frame 1700\AA\ very closely
matches the observed-frame \I-band at \zs4, \R-band at \zs3, and
\G-band at \zs1.7, thereby nearly eliminating uncertainties 
in $k$-corrections.  At \zs2.2, 1700\AA\ is located between the \G\
and \R\ bands and so for our \zs2.2 objects we construct composite
\GRave\ magnitudes that more closely match rest-frame 1700\AA\ than do
either \G-band or \R-band alone.  Our composite \GRave\ magnitudes are
a simple average of the \G-band and \R-band fluxes,
\begin{equation}\label{defGRave.eq} 
G{\cal R} = -2.5\log[(10^{-0.4G}+ 10^{-0.4{\cal R}})/2], 
\end{equation}
with uncertainties calculated by combining the \G-band and \R-band
uncertainties in quadrature.

\subsection{Calculating the effective volumes of the survey}\label{Veffcalc}

We must first compute the effective volume of the survey, as a
function of apparent magnitude, for objects in each redshift sample.
Because galaxies scatter in and out of the color-color selection
regions (see Fig.~\ref{colorcolor.fig} here and Figs.~4 and 5 in
\kdfdata) that we use to select our high-$z$ galaxy samples, and
because this scattering will depend on the size of photometric
uncertainties, we must compute our effective volumes as a function of
apparent magnitude.  Our calculation of effective volumes is
accomplished through simulations that we implant into our images and
then seek to recover artificial objects with colors and magnitudes
representative of star-forming high-$z$ galaxies.  The description of
these simulations is the subject of the present section,
\S~\ref{Veffcalc}.

\subsubsection{Modeling the colors of high-$z$ galaxies}\label{modelcolors}

As the first step, we calculate a grid of model colors expected of
high-$z$ galaxies.  We start with model spectral energy distributions
(SEDs) of star-forming galaxies from the 1996 version of the Bruzual
\& Charlot (1993) spectral synthesis library.  We use the continually 
star-forming models with solar metallicity and Salpeter initial mass
function (IMF).  We next redden them with a set of extinction values
in the the Calzetti (1997) starburst dust prescription.  We then
complete the redshift dimension of the grid by stretching these
SEDs by $(1+z)$ and attenuating them using the Madau (1995)
prescription for continuum and line blanketing due to intergalactic
hydrogen along the line of sight.  Finally, we integrate the resultant
reddened, observer-frame model spectra through the
\UGRI\ filter transmission curves to arrive at the predicted colors of
high-$z$ star-forming galaxies.  Some examples of model galaxy colors
are shown in Fig.~\ref{colorcolor.fig}, where they are overplotted on
top of regions of color-color space used to define our galaxy samples.

The largest influence on the colors of high-$z$ galaxies is wielded
by, first, attenuation by intergalactic hydrogen gas blueward of the
Lyman break, and, second, by reddening due to interstellar dust
internal to the galaxies.  Other variables such as age, star formation
history, stellar initial mass function, or metallicity can also play a
role, but their effects are small in comparison, and especially so at
the rest-UV wavelenghts that concern us in this study.  Consequently,
we adopt fixed values for most of these parameters and explore only
how our results vary with the adopted reddening and starburst age.

For practical reasons, we must restrict our choices to a limited set
of these parameters.  We are guided in our choice of dust attenuation
and starburst age by the observed values of these quantities in \zs3
LBGs.  Early on, Sawicki \& Yee (1998) studied the rest-frame UV
through optical broadband photometry of 17 spectroscopically-confirmed
\zs3 LBGs in the HDF and concluded that these objects are dominated by
young stellar populations ($\lesssim$0.2 Gyr) and substantial amounts
of dust (median
\ebv$\sim$0.3).  However, the bulk (11/17) of the objects in their
analysis came from the spectroscopic sample of Lowenthal et al.\
(1997) who allowed objects that are redder --- and so presumably more
dusty --- than those selected using the now so familiar criteria of
Steidel et al.\ that are used in our present KDF work.  Indeed,
Shapley et al.\ (2001) applied the SED-fitting technique of Sawicki \&
Yee (1998) to a large sample of \zs3 LBGs selected {\it solely} using
the Steidel et al.\ selection criteria, and found a median \ebv=0.16
(lower than Sawicki \& Yee 1998) and concluded that LBGs undergo
relatively short periods (50--100 Myr) of very intense star formation
followed by a more quiescent star-forming phase (see also Sawicki \&
Yee 1998).

It is thus clear that at least bright (\R$\lesssim$25) LBGs at
\zs3 are dominated by fairly short episodes of star formation and 
significant dust obscuration.  It remains unclear whether this is also
the case at lower and higher redshifts and at the fainter magnitudes
that we reach in the KDF.  Nevertheless, motivated by the results of
Sawicki \& Yee (1998) and Shapley et al.\ (2001), we take as our
fiducial model the 100~Myr-old star-forming SED from the 1996 version
of the Bruzual \& Charlot (1993) and attenuate it with \ebv=0.15 of
dust.  In \S~\ref{p} we show that this fiducial model reproduces the
Steidel et al.\ (1999, 2003, 2004) observed redshift distributions of
\zs2.2, \zs3, and \zs4 galaxy samples (the case for \zs1.7 is less clear).  
Nevertheless, to monitor the impact of our choice of SED model on our
LF results we carry out all our calculations in parallel, considering
a grid of SED models that includes two stellar population ages, 10~Myr
and 100~Myr, and seven values of dust attenuation, \ebv=0--0.3 in
steps of 0.05.  As we will discuss in \S~\ref{systematics}, the
dependence of the LF on these assumed dust and age values is
negligibly small at \zs4 and \zs3, but becomes more significant at the
lower redshifts.

\subsubsection{The sample completeness function \pmz}\label{p}

Next, we must correct for incompleteness of our catalogs that is
brought on both by the imperfect object detection efficiency and by
the scatter of high-$z$ galaxies across the boundaries of our
color-color selection boxes.  We do not (here or elsewhere in our LF
calculation) explicitly correct for foreground, low-$z$ interlopers
that contaminate our high-$z$ sample.  Such interloper contamination
is known to be very small in the spectroscopic samples of Steidel et
al.\ (1999, 2003, 2004).  Because --- as is discussed in \kdfdata\ ---
the contamination fraction is expected to at worst remain constant and
in all likelihood to fall towards fainter magnitudes, the
contamination should be equally small or even smaller in our KDF data.

We measure the amount of incompleteness by implanting simulated
galaxies into our images and then seeking to recover them using the
very same procedures that we used in making our data catalogs (see
\kdfdata).  Incompleteness (both detection incompleteness and the loss
of galaxies due to scattering out of the high-$z$ color-color
selection boxes) is a function of apparent magnitude, with fainter
galaxies suffering larger incompleteness than their brighter kin.  It
also depends on the true colors of our target galaxies, and so their
redshifts and intrinsic SEDs.  We insert artificial objects with the
expected colors of high-$z$ galaxies generated as described in
\S~\ref{modelcolors}.  Our artificial objects have point-source 
profiles because --- as we discussed in \kdfdata --- the seeing in our
images is sufficiently poor (FWHM$\sim$1\arcsec) to ensure that
high-$z$ galaxies are spatially unresolved.  The recovered fractions
form the completeness function \pmz, which is the probability that a
galaxy of a given apparent magnitude (in \I\ at \zs~4, \R\ at \zs3,
\GRave\ at \zs2.2, and \G\ at \zs1.7) and redshift $z$, dust attenuation,
and model age, matches our sample selection criteria. 

The function \pmz\ is measured separately for each of the four
redshift samples (\zs4, \zs3, \zs2.2, and \zs1.7) and --- given the
small differences in the image properties of our five KDF fields ---
is recalculated for each KDF field.  The function is sampled in steps
of $\Delta m$=0.5 in input apparent magnitude and $\Delta z$=0.1 in
redshift, and for the 8 combinations of age and reddening discussed
above.  At each step in this parameter grid several hundred simulated
objects are implanted at quasi-random locations in the image.  These
positions are always the same for the different steps in the parameter
grid, but are otherwise unremarkable and sample the images fairly.

We can use our calculated \pmz\ to test whether our assumptions about
age and amount of dust are reasonable --- i.e., whether we can
reproduce the observed redshift distributions of high-$z$ populations.
The shaded histograms in Fig.~\ref{zdistr.fig} show the redshift
distributions of the spectroscopic samples of Steidel et al.\ (1999,
2003, 2004).  These Steidel et al.\ spectroscopic samples contain
galaxies with a range of apparent magnitudes, and represent the
underlying color-selected population convolved with a spectroscopic
success function that is not trivial to model.  However, the bulk of
their spectroscopic samples consists of galaxies with \R~$\sim$25 ---
i.e., galaxies with photometric errors that are similar to those of
\R~$\sim$26.5 galaxies in the KDF.

Thus, our \pmz\ for \R=26.5 KDF galaxies should match the observed
redshift distributions if our modeling is a reasonable representation
of reality.  Fig~\ref{zdistr.fig} shows that this is indeed the case
for our fiducial model with \ebv=0.15 and starburst age of 100 Myr
(the 10 Myr models, which are not shown, are also very good): Within
each panel of Fig.~\ref{zdistr.fig}, the solid curves show the
function $p($\R=26.5$,z)$ for the three values of \ebv\ (which
generally increase from left to right), with the thicker curve marking
\ebv=0.15.  Although our fiducial model does not give a
uniquely-matching solution (for example, a superposition of lower and
higher \ebv\ values might work just as well), it does give {\it a}
good agreement with the data.  The fiducial model works remarkably
well at \zs4, 3, and 2.2; it works less well at \zs1.7 as it predicts
a redshift distribution with a somewhat lower median redshift than is
observed, but so do all our other \pmz\ at that redshift (we will
revisit this issue later).  At any rate, the ability of our \pmz\
modeling to reproduce the observed redshift distributions gives us
confidence in that modeling, in the modeling of effective survey
volumes, \Veff, that are based on it in the next subsection, and
thence in our estimate of the luminosity functions.

\subsubsection{The effective volumes}

Finally, \Veff\ is calculated for each of the five fields by
integrating the probability function \pmz\ over redshift:
\begin{equation}
V_{eff}(m) = A_f \int \frac{dV}{dz} p(m,z) dz,
\end{equation}
where $dV/dz$ is the comoving volume per square arcminute in redshift
slice $dz$ at redshift $z$ and $A_f$ is the area of the field in
arcminutes.  The \Veff\ is calculated separately for each
color-selected redshift sample (\zs4,
\zs3, \zs2.2, \zs1.7), each galaxy model (which consist of the eight 
combinations of reddening and starburst age), each apparent magnitude
step ($\Delta m$=0.5 in \I, \R, \GRave, or \G, depending on the
redshift sample being considered), and each of the five KDF fields.

\subsection{The incompleteness-corrected number counts}\label{numbercounts}

We next calculate the in\-com\-ple\-te\-ness-corrected galaxy density
$\phi(m)$ as a function of apparent magnitude and for each \Veff\
model.  We carry out the calculation separately for each of the
redshift samples, \zs4, \zs3, \zs2.2, and \zs1.7.  The calculation is
first carried out independently for each KDF field and then the
results are averaged together.

For each field $f$ we compute in 0.5-mag bins the number of galaxies
satisfying the color selection criteria, and then correct for
incompleteness using the effective volume:
\begin{equation}\label{phim.eq}
\phi_{f}(m)=2\frac{N_f(m)}{V_{eff}(m)}.
\end{equation}
Here, $N_f(m)$ is the number of observed galaxies within the magnitude
bin $m\pm0.25$ in that field, \Veff(m) is the effective volume of that
field, $\phi_{f}(m)$ are the incompleteness corrected number counts in
units of mag$^{-1}$Mpc$^{-3}$ and the factor of 2 converts from counts
in the 0.5 mag bins to counts per mag.

The results of the individual fields are then weighted by field area,
$A_f$, and averaged to yield the in\-com\-ple\-te\-ness-corrected
galaxy number density for the entire KDF survey,
\begin{equation}\label{phimALL.eq}
\phi(m) = \frac{\Sigma_{f}A_{f}\phi_{f}(m)}{\Sigma_{f}A_{f}}.
\end{equation}
We restrict the calculation of the final KDF $\phi(m)$ to magnitude
bins no fainter than the 50\% detection completeness limit in each
field of the KDF.  This limit is deeper by 0.5 mag for three of our
fields than for the other two (see \S~\ref{introKDF} and \kdfdata) and
so the average $\phi(m)$ in Eq.~\ref{phimALL.eq} is computed using all
five fields at all magnitudes except for the faintest magnitude bin in
which only the deeper three fields are used.

The uncertainty in $\phi(m)$ combines two sources of uncertainty,
namely the uncertainty in number statistics, $\delta\phi_N(m)$, and an
estimate of field-to-field fluctuations, $\delta\phi_{f2f}(m)$. The
number statistics uncertainty is simply the Gaussian $\delta\phi_N(m)
= \phi(m)[N(m)]^{-0.5}$, where N(m) is the total number of galaxies in
all five (or three) fields in that magnitude bin.  The field-to-field
uncertainty $\delta\phi_{f2f}$ is estimated using bootstrap
resampling, whereby we generate 500 new realizations of $\phi(m)$ via
Eq.~\ref{phimALL.eq}, but now in each realization choosing fields
randomly with replacement and so allowing the same field to be
included more than once (or not at all) in a given realization.  The
field-to-field uncertainty $\delta\phi_{f2f}$ is then taken to be the
rms value of the 500 $\phi(m)$ resampled realizations.  These two
sources of uncertainty are then added in quadrature $\delta\phi =
[(\delta\phi_N)^2 + (\delta\phi_{f2f})^2]^{0.5}$ to give us the total
uncertainty.

\subsection{Absolute magnitudes}\label{absmags}

The final step is to convert the \phim\ into the luminosity function
\phiM.  As we discussed earlier, we compute the luminosity function at 
rest-frame 1700\AA\ both to minimize $k$-corrections, and to retain
commonality with the work of Steidel et al.\ (1999) at brighter
magnitudes.

The absolute magnitude, $M$, is derived using the usual cosmological
distance modulus, $DM$, and $k$-correction, $K$,
\begin{equation}
M_{1700} = m_{\lambda_{obs}} - DM - K,
\end{equation}
which we rewrite as 
\begin{eqnarray}\label{m2M.eq}
M_{1700} & = & m_{\lambda_{obs}} - 5 \log (D_L / 10pc) + 2.5 \log (1+z) \nonumber \\
         &   & + (m_{1700}- m_{\lambda_{obs}/(1+z)}).
\end{eqnarray}
Here, $D_L$ is the luminosity distance and $m_{\lambda_{obs}}$ is the
observed magnitude in the principal filter for the redshift sample
being considered (\I, \R, \GRave, and \G\ for \zs4, 3, 2.2, and 1.7,
respectively).  The last term of Eq.~\ref{m2M.eq},
$(m_{1700}-m_{\lambda_{obs}/(1+z)})$, is the $k$-correction color
between rest-frame 1700\AA\ and the principal filter in the {\it
rest-frame} for the redshift sample in question.  This $k$-correction
color is expected to be very small because of our decision to work at
rest-frame 1700\AA.  This expectation is illustrated in the top two
panels of Fig.~\ref{kcorr.fig}, where we plot the $k$-correction color
for two representative galaxy models selected out of a larger ensemble
that we tested.  As Fig.~\ref{kcorr.fig} illustrates, the value of the
color term is indeed very close to zero for \I-band at \zs4, \R-band
at \zs3, the composite \GRave\ at \zs2.2, and \G-band at \zs1.7.  Over
the redshift ranges selected by the color selection criteria we are
using, the deviations from zero are typically no larger than 0.1~mag
with a redshift-dependent $\delta$mag {\it range} of no more than
$\sim$$\pm$0.1 for a given model.  These small offsets are negligible
and so we set $(m_{1700} - m_{\lambda_{obs}/(1+z)})$ to zerßo in
Eq.~\ref{m2M.eq} to arrive at
\begin{equation}\label{absmag.eq}
M_{1700} = m_{\lambda_{obs}} - 5 \log (D_L / 10pc) + 2.5 \log (1+z).
\end{equation}
Applying Eq.~\ref{absmag.eq} to our \phim\ we at last arrive at the
rest-frame 1700\AA\ luminosity function, \phiM.  The resulting
luminosity function measurements are described in the following
section, \S~\ref{theLF}.

\section{THE OBSERVED LUMINOSITY FUNCTIONS}\label{theLF}

\subsection{Description of the luminosity functions}

The data points in Fig.~\ref{lf4pan.fig} show our LF measured using
our baseline \Veff\ model of 100~Myr-old starburst with \ebv=0.15
(dependence on model assumptions will be discussed in
\S~\ref{systematics}).  In addition to the KDF data we also include 
the \zs4 and \zs3 LF points of Steidel et al.\ (1999).  We stress that
our analysis of the KDF data follows closely the selection and LF
analysis procedures used by Steidel et al.\ (1999) for these brighter
galaxies and so combining their results with ours should be robust and
free of systematic uncertainties.  The KDF and the Steidel et al.\
samples complement each other: the Steidel et al.\ (1999) measurements
provide good statistics at the bright end, but do not probe fainter
than \Muv=$-$20.5 at \zs3 and \Muv=$-$21 at \zs4, while the KDF have
good statistics at the faint end, probing $\sim$1.5 magnitudes deeper,
but lack the statistics at the bright end.  Unless otherwise stated,
throughout the rest of this paper we will always combine the Steidel
et al.\ (1999) measurements and our fainter KDF data when discussing
the \zs3 and \zs4 LFs.

To give a more physical meaning to our luminosity scale we relate
rest-frame UV magnitudes to star formation rates.  Combining the
relation between SFR and luminosity given by Kennicutt (1998; see also
Madau, Pozzetti, \& Dickinson 1998) with the definition of AB
magnitudes (Oke 1974) gives
\begin {equation}\label{muv2sfr.eq}
SFR = 6.1 \times 10^{ -(8+0.4M_{1700})} M_\odot yr^{-1}.
\end{equation}
This conversion is valid in the absence of dust and for a stellar
population that is forming stars continuously and whose UV light is
dominated by massive, short-lived stars that are being produced on a
Salpeter (1955) stellar initial mass function (IMF) with mass range
0.1$\leq$\Msun$\leq$100.  The SFR scales are plotted on the top axes
in Fig.~\ref{lf4pan.fig}.  The SFR scale should not be taken too
literally because of the uncertainties in the assumptions underlying
Eq.~\ref{muv2sfr.eq}.  Nevertheless, this scale gives us a useful,
more physical reference frame for the luminosity functions.  As
Fig.~\ref{lf4pan.fig} shows, the KDF LF reaches to
\Muv=$-$18 at \zs2.2 and \Muv=$-$19 at \zs4.  It thus rivals in depth 
LFs measured in the HDFs (e.g., Sawicki, Lin, Yee 1997; Steidel et
al.\ 1999).  Under the aforementioned assumptions, our KDF LF reaches
down to galaxies with star formation rates of $\sim$2 \Msun/yr and
lower --- comparable to the SFR in the Milky Way today.

The solid curves in each panel of Fig.~\ref{lf4pan.fig} show Schechter
function fits to the data (Schechter 1976), and the gray shaded
regions show the 1$\sigma$ uncertainties in those fits. We defer the
description of the details of the Schechter function fitting to
\S~\ref{schechterfits} and here we use the fits only to guide the eye
and to give a first comparison between LFs at different redshifts.

Our \zs3 LF is the best constrained of all the redshift bins we
consider and so, when comparing luminosity functions at different
redshifts, we will use the \zs3 LF as reference.  The dashed curves in
the other panels of Fig~\ref{lf4pan.fig} show the fit to this \zs3
fiducial.  Comparing the data at \zs4 with our fiducial
\zs3 curves immediately suggests that the LF undergoes evolution with
redshift: The number density of faint galaxies at \zs4 appears to be
significantly lower than at \zs3.  At the same time, the number
density of luminous galaxies appears to remain unchanged from \zs4 to
\zs3.  In \S\S~\ref{systematics} and \S~\ref{LFevolution} we will explore 
in detail whether this LF evolution is real or simply the result of a
selection effect or other artefacts (we conclude that it is very
likely real at least between \zs4 and \zs3).

For comparison, Fig.~\ref{lf4pan.fig} also shows the recent low-$z$
rest-1500\AA\ LF measurements from the GALEX mission.  These LFs are
shown as dotted curves in Fig.~\ref{lf4pan.fig}, with the left dotted
curve showing the \zs1 GALEX LF and right one showing the
\zs0 LF (Arnouts et al.\ 2005 and Wyder et al.\ 2005, respectively).
The strong evolution of the LF of star-forming galaxies from \zs0 to
\zs1 seen in the GALEX data has been recognized for some time 
(e.g., Lilly et al.\ 1995) and is responsible for the steep rise in
the UV comoving luminosity density of the Universe over that redshift
interval (e.g., Lilly et al.\ 1996; Schiminovich et al. 2004).
Similarly, the increase in the number of luminous galaxies from \zs1
to higher redshifts, $z$$\gtrsim2$ is part of the familiar UV-sketch
of the cosmic star formation history (e.g., Madau et al.\ 1996;
Sawicki, Lin, \& Yee 1997; Giavalisco et al.\ 2004).  In contrast to
these well-known broad trends, the more subtle evolution in the LF
from \zs4 through \zs3 to \zs2.2 that is revealed in our KDF data has
not been so far explored because until now there was a lack of
well-selected and well-tested samples that have good statistics over a
wide range in luminosity.

\subsection{Parametric representation of the LF}\label{schechterfits}

We fit the binned luminosity function data with the Schechter (1976)
function,
\begin{eqnarray}\label{schechterfunction.eq}
\phi_{model}(M) & = & \phi^*\hat{\phi}(M)\\
	        & = & \phi^*0.4ln(10) dex\{[0.4(M^*-M)]^{(1+\alpha)}\}\nonumber\\
		&   & \times exp[-10^{0.4(M^*-M)}]\nonumber,
\end{eqnarray}
which we evaluate over \Mstar, \phistar, and $\alpha$.  In practice,
we do the fitting using \chisq\ minimization, where for a grid of
\Mstar\ and $\alpha$ values we compute a corresponding grid of \chisq\
values using
\begin{equation}\label{schechterchisq.eq}
\chi^2(M^*,\phi^*,\alpha)=\sum_M\Big[\frac{\phi_{data}(M)-\phi_{model}(M)}{\sigma(M)}\Big]^2,
\end{equation}
where the sum is taken over the \Muv\ magnitude bins, and
$\phi_{model}(M)$ are computed using Eq.~\ref{schechterfunction.eq}.
Instead of adding a third, \phistar, dimension to the grid, the
computation is considerably accelerated by optimally calculating
\phistar\ at each (\Mstar, $\alpha$) in the grid using the analytic
relation 
\begin{equation}\label{phistar.eq}
\phi^*=\frac{\sum_M \hat{\phi}(M)\phi_{data}(M)/\sigma^2(M)}{\sum_M\hat{\phi}^2(M)/\sigma^2(M)},
\end{equation}
which is derived by minimizing Eq.~\ref{schechterchisq.eq} via
$\partial\chi^2/\partial\phi^*=0$.  We then search the grid of \chisq\
values to select its minimum, $\chi_{min}^2$, and adopt its
corresponding \Mstar, \phistar, and $\alpha$ as the best-fit Schechter
function parameters.  The values of these best-fit parameters are
listed in Table~\ref{schechter.tab} and plotted in
Fig.~\ref{schechter.fig}.

The error contours shown in Fig.~\ref{schechter.fig} are computed by
recalculating the best-fitting \Mstar, \phistar, and $\alpha$, but now
with sets of data values $\phi_{data}(M)$ that have been perturbed
randomly according to their standard deviations $\sigma(M)$. For each
of the redshift bins, we generate 250 such perturbed realizations and
use their \chisq\ to map out the regions of parameter space that
correspond to the best-fitting 68.3\% of such realizations.

\subsection{Systematic effects in the LF measurement}\label{systematics}

\subsubsection{Uncertainty due to $k$-corrections}

As we discussed in \S~\ref{absmags}, setting the $k$-correction color
term to zero introduces only a very small, $\lesssim$0.1 mag,
systematic bias in the determination of absolute magnitude (see
Fig.~\ref{kcorr.fig}).  Consequently, $\sim$0.1 mag can be taken as
the systematic uncertainty in our determination of the positions of
\phiM\ bins.  This uncertainty is too small to affect our LFs significantly 
and so we do not consider it further.

The lack of spectroscopic redshift information for objects in our
sample may also introduce a systematic effect, albeit --- as we show
here --- a negligible one.  Our lack of spectroscopic redshifts means
that we do not know whether a particular object in a given
color-selected redshift sample is near the lower or the higher end of
the redshift interval.  This uncertainty may introduce a systematic
bias since at the same apparent magnitude intrinsically less luminous
objects are more likely to come from somewhat lower redshifts (and
hence occupy a smaller effective volume) than intrinsically more
luminous galaxies.  However, as is shown in the bottom panels of
Fig.~\ref{kcorr.fig}, the effect of this redshift uncertainty on the
value of the distance modulus $DM$ and hence on the derived $M_{1700}$
is small: it is $\pm$0.2~mag for the \zs4 and \zs3 samples,
$\pm$0.4~mag at lower redshifts.  Such systematic offsets of a few
tenths of a magnitude in \Muv\ are comparable to uncertainties
introduced into the LF measurement by \rootN\ statistics and
field-to-field variance as reflected, for example, in the
uncertainties shown in Fig.~\ref{schechter.fig}.  Moreover, because
the shapes of the LFs are similar in all redshift bins, the systematic
bias introduced by this effect works in the same direction at \zs4, 3,
2.2, and 1.7, thereby reducing any systematic differences {\it
between} the redshift bins.  In summary, then, the biases introduced
by our lack of spectroscopic redshifts are highly unlikely to
drastically affect our results.

\subsubsection{Dependence on cosmology}\label{dependence-on-cosmology}

Throughout this paper we have assumed a cosmological model with
\OmOlHo=(\cosmoparams).  However, it is well known that many
quantities, including the luminosities and volume elements directly
relevant in LF calculation, can depend strongly on the assumed
cosmological model.  Consequently, we feel it important to explore how
the choice of cosmological model impacts our results.

The choice of cosmology affects the derived luminosity function
primarily through a change in the relative number density
normalization (i.e., a change in $\phi$; Eq.~\ref{phim.eq}) and
luminosity (i.e., a change in the absolute magnitude scale, $M$;
Eq.~\ref{absmag.eq}).  The dependence of the faint-end slope is weak
as it enters only through a small difference in how the change in
cosmology affects the derived \Veff\ at different apparent magnitudes.
We have tested that the effect on the faint-end slope is negligible by
recomputing our LFs for a model with \OmOlHo=(1, 0, 100 km s$^{-1}$
Mpc$^{-1}$) and consequently, we will focus our discussion on the
effects on $M$ and $\phi$.

Figure~\ref{comparecosmo.fig} shows how changing cosmology from our
assumed \OmOlHo=(\cosmoparams) to two other often-used cosmological
models changes the resultant absolute magnitudes (top panel) and
number densities (bottom panel).  In both cases, the effect can be
quite large: the difference in absolute magnitudes between the two
models can be as large as 1--2 mags, and the change in number density
as large as an order of magnitude.  However, the {\it relative} change
in $M$ and $\phi$ are only weakly dependent on redshift for a given
cosmological model over the redshift range we study: {\it relative to}
\zs3, the \zs1.7, \zs2.2 and \zs4 $M$ scales are shifted by less than
$\sim$0.15 mag by a change of our assumed cosmology; similarly, the
number densities are also affected only weakly, at the $\lesssim$20\%
level.  A direct consequence of this is that while the {\it absolute}
$\phi$ and $M$ scales depend strongly on cosmology, the {\it relative}
shapes and $\phi$ and $M$ normalizations from redshift bin to redshift
bin are virtually unchanged.  As a result, {\it any real evolutionary
trends seen in the LF from \zs4 to \zs3 to \zs2.2 to
\zs1.7 are virtually independent of the assumed cosmology.}

\subsubsection{Dependence on galaxy model properties}
\label{dependence-on-model-properties}

Our LF calculation relies --- as it must --- on our estimates of the
effective volumes of the survey (see \S~\ref{LFcalculation}).  These
\Veff\ estimates are based on models of the expected colors of
high-$z$ galaxies and therefore may depend critically on our model
assumptions.  Throughout this paper we have assumed a baseline model
for high-$z$ galaxies based on a 100-Myr old starburst reddened by
\ebv=0.15 of dust.  As we have shown in \S~\ref{p}, this model reproduces 
well the observed redshift distributions of high-$z$ galaxies, while
models that are similar but have different amounts of dust do a poorer
job.  Nevertheless, we wish to explore how strongly model-dependent
are our \Veff\ estimates and the resultant LF measurements.

To test how model-dependent our LF results are, we have repeated all
the steps in our LF calculation (\S~\ref{Veffcalc} ---
\S~\ref{schechterfits}) for a grid of galaxy models that contains
starbursts of two different ages (10 and 100 Myr) and eight values of
\ebv\ (0--0.3 in steps of 0.05, and also a composite model that 
contains a mixture of galaxies with \ebv\ values drawn evenly from
that \ebv\ range).  We note that we have applied the different models
to our own KDF points but were unable to do so to the points that come
from Steidel et al.\ 1999 since we do not know in detail the \Veff\
model that was used in their LF work.  Consequently, the effect of
changing \Veff\ models will be seen only in our data and so will
manifest itself fully at the faint end of the LFs, while the bright
ends of the \zs4 and \zs3 LFs, which are dominated by the Steidel et
al.\ data, will remain unchanged.

We show the effect of changing the \Veff\ model in two ways.  First,
the recomputed LFs are plotted in Fig.~\ref{LF-4panel-modelcomp.fig},
in which the data points at each \Muv\ show the results of 7 LFs
computed using the seven discrete \ebv\ values, and the solid black
curves show all the corresponding Schechter function fits; the gray
curves in the \zs4, \zs2.2, and \zs1.7 panels reproduce the \zs3 LF
Schechter function fits for comparison.  Second, the number density
value of the Schechter function fits at a fixed magnitude, \Muv=$-$20,
are shown as a function of \ebv\ in Fig.~\ref{fixedphi-vs-model.fig};
here, the black lines connect the 100 Myr-old starburst models and the
lighter lines are for the 10 Myr-old ones.

As can be seen in the \zs3 panel of
Fig.~\ref{LF-4panel-modelcomp.fig}, the choice of \Veff\ galaxy model
has a miniscule effect at \zs3 --- an effect of changing the \Veff\
model has an effect that is no larger than the errors due to number
statistics and cosmic variance (see Fig~\ref{lf4pan.fig}).
Fig.~\ref{fixedphi-vs-model.fig} confirms that at \zs3 the dependence
on model is small: the number density of galaxies at \Muv=$-$20 ranges
over the range of models by $\lesssim$5\% compared to our baseline
model.  We therefore conclude that at \zs3 the LF we measure is very
robust with respect to model assumptions.

While the results at \zs3 are virtually model-independent, the choice
of \Veff\ model has a larger (although still small) effect at
\zs4.  Here, the  $\phi_{M=-20}$ values range by up to $\pm$$\sim$20\% 
with respect to our baseline model.  However, despite this variation,
the faint end of the \zs4 LF {\it always} remains substantially below
the \zs3 LF, and the variation between different models at \zs4 is not
larger than the uncertainties introduced by field-to-field variations
and \rootN\ statistics, as can be seen by comparing
Figs.~\ref{LF-4panel-modelcomp.fig} and \ref{lf4pan.fig}.  We thus
conclude that while the choice of \Veff\ model may possibly affect the
{\it details} of the \zs4 LF results, it is unlikely to alter the
qualitative trends seen between the \zs4 and \zs3 LFs.

The choice of \Veff\ potentially has the largest effect at \zs2.2.
While models with \ebv$>$0 give a tight grouping of LF results, the
two models with \ebv=0 give LFs that have a substantially higher
$\phi$ normalization.  These higher LFs are a direct result of the
fact that the colors of \zs2.2 models with \ebv=0 approach very
closely to the boundary of the color selection box (see Figs.~4 and 5
in \kdfdata): large numbers of model galaxies in these models are
scattered out of the color selection box, resulting in low \pmz\
values, low \Veff, and hence high number densities $\phi$ in the
resultant LFs.  In our case the effect is not critical because (as we
have argued in \S~\ref{p}) the \ebv=0 models are probably not
realistic on other grounds; moreover, even the adoption of the extreme
\ebv=0 results does negate the possibility that the evolution seen 
from \zs4 to \zs3 continues onwards to \zs2.2 --- in fact, an \ebv=0
makes such an evolutionary trend stronger at \zs2.2 than in the case
of other \ebv values.  However, our \ebv=0 case at \zs2.2 illustrates
that in general one must be careful when computing luminosity
functions of color-selected samples because such LFs can be very
strongly dependent on the {\it assumed} properties of the high-$z$
galaxy populations that one is trying to study.  

At \zs1.7 the LF appears to have an even higher number density
normalization than at higher redshifts.  We note however that we do
not place much faith in our determination of the \zs1.7 LF as we
discuss in more detail in \S~\ref{z17evolution}.

In summary, we conclude that while the choice of \Veff\ models used in
calculating the LF may have an effect on the LF results, {\it in our
case} such effects are small: they are no larger than the
uncertainties introduced by field-to-field variance and by $\sqrt{N}$
statistics and in any case do not affect the qualitative evolutionary
trends seen between \zs4, \zs3.  Meanwhile, the \zs2.2 model \Veff\
may suffer from systematics that may adversely affect the accuracy of
our LF measurements at those redshifts.

\subsubsection{Field-to-field fluctuations}\label{field2field}

Galaxy clustering introduces field-to-field fluctuations in the galaxy
distribution.  This effect --- often termed cosmic
variance\footnote{This name is not strictly correct as cosmic variance
refers to the variance in samples that are {\it fundamentally} limited
by the finite size of the Universe, such as, e.g., the largest-scale
fluctuations on the Cosmic Microwave Background.  Nevertheless,
because it has entered the common nomenclature of galaxy evolution
studies, we use the term cosmic variance interchangably with
field-to-field variance throughout this paper.} --- limits the
accuracy with which the luminosity function can be measured.  This is
particularly true for surveys that rely on small single pointings,
such as the HDF (e.g., Sawicki et al.\ 1997; Steidel et al.\ 1999),
FDF (Gabasch et al.\ 2004) or the Hubble Ultra Deep Field (e.g.,
Bouwens et al.\ 2006).

To mitigate the effects of cosmic variance, the KDF consists of five
{\it large} fields grouped in three spatially-independent patches.
When calculating the luminosity function we have taken field-to-field
variance into account by including a bootstrap resampling measurement
of this variance in our errorbars (see \S~\ref{numbercounts}).  Here,
we illustrate the strength of field-to-field fluctuations by computing
the characteristic galaxy number density \phistar\ in each of the five
KDF fields.  We make the measurement of the \phistar\ while holding
the shape of the luminosity function fixed.  Specifically, at each
redshift we hold $\alpha$ and \Mstar\ constant at the values we
measured earlier for the full dataset (Table~\ref{schechter.tab})
while letting \phistar\ be a free parameter.  We also exclude the
Steidel et al.\ (1999) \zs4 and \zs3 bright-end data from the
measurement here.  Table~\ref{phistar_by_field.tab} summarizes the
results and contrasts them with the \phistar\ values for the full
dataset.  The field-to-field fluctuations are generally small.  They
are largest at \zs4 where the total number of objects in our sample is
smallest: here, the largest excursion --- that in the 03B field --- is
$\sim$1.5 times the fiducial value, although the RMS scatter is
significantly smaller than that and consistent with our bootstrap
estimates.

These field-to-field fluctuations, while relatively small in the KDF,
underscore the need for multiple sightlines when determining the
galaxy number density, luminosity function, and derived quantities
such as luminosity and star formation rate densities of the universe.
Single-field studies, especially if they rely on {\it small} fields
such as the HDF or the UDF, have no way of monitoring this important
source of error.  In contrast, here, in the multi-field KDF, we have
estimated field-to-field variance throught bootstrap resampling and
included it explicitly in our error budget.

\subsection{Comparison with other surveys}\label{othersurveys}

Over the last few years several authors have presented luminosity
functions of UV-selected galaxies at high redshift, $z$$>$1.  The vast
majority of these measurements were made using either full-blown
photometric redshifts or the two-color selection techniques inspired
by the success of the Steidel et al.\ surveys.  The bulk of this work
can be divided into two groups: those that use the very deep images of
one or both Hubble Deep Fields (HDFs; Williams et al.\ 1996; Casertano
et al.\ 2002) to probe the LF to very faint limits that are comparable
to the KDF (e.g., Gwyn \& Hartwick 1996; Sawicki et al.\ 1997; Steidel
et al.\ 1999, ), and those that use wider but shallower ground-based
data (e.g., Steidel et al.\ 1999; Iwata et al.\ 2003; Ouchi et al.\
2004a; Gabasch et al.\ 2004).

Here, we compare our \zs3 and \zs4 LFs with two recent wide but deep
surveys, namely the \zs4 LF of the Subaru Deep Survey (SDS; Ouchi et
al.\ 2004a) and with the \zs3 and \zs4 LFs of the FORS Deep Field
(FDF; Gabasch et al. 2004).  We also compare our measurement of the
LF's faint end with the Steidel et al.\ (1999) analysis of the
HDF-North, for although the HDFs are too small to be truly adequate
for LF determination, the Steidel et al.\ (1999) HDF-based faint-end
slope of $\alpha$=$-$1.6 is often used in the literature, particularly
by workers deriving the integrated UV luminosity density and star
formation density of the Universe at higher redshifts (e.g., Bouwens
et al.\ 2004; Bunker et al.\ 2004; Dickinson et al.\ 2004; Giavalisco
et al.\ 2004; Yan \& Windhorst 2004).

\subsubsection{The Hubble Deep Field LF of Steidel et al.\ (1999)}

Steidel et al. (1999) have applied a modified version of their
color-color LBG selection criteria to the northern Hubble Deep Field
in order to extend to fainter magnitudes their \zs3 and \zs4 LF
analysis.  Their \zs3 and \zs4 HDF LFs are shown in
Fig.~\ref{fig-LF-otherpple.fig} as upward-pointing triangles.  At
\zs3, our KDF LF (filled circles) is in good qualitative agreement with 
the Steidel et al.\ 1999 HDF results albeit the KDF --- with its much
larger area and multiple pointings --- provides a much more robust and
precise measurement.  Our faint-end slope
$\alpha$=$-$1.43$^{+0.17}_{-0.09}$ is significantly shallower than the
$\alpha$=$-$1.6 reported by Steidel et al.\ from the HDF (or that by
Sawicki et al.\ 1997 who measured the high-$z$ LFs in the HDF using a
different selection technique).  As we will discuss in Sawicki \&
Thompson (in prep., \kdfsfhist), the consequences of this difference
in $\alpha$ are not negligible in calculations of the UV luminosity
density and star formation density of the Universe.  At \zs4, the HDF
data suffer from very poor statistics in addition to potential
problems with cosmic variance, but they are lower than the \zs3 points
and, within their large errorbars, agree well with the KDF \zs4 LF.

\subsubsection{The FORS Deep Field}

The FORS Deep Field reaches depths similar to those of the KDF in a
single VLT pointing that covers $\sim$50~\sqam, or 29\% the area of
the KDF. The fact that it comprises only a single field means that it
is highly subject to cosmic variance effects, but --- unlike the HDFs
--- it is sufficiently large that it is not dominated by small number
statistics at the faint end.  Gabasch et al.\ (2004) have computed
photometric redshifts in the FDF and used them to estimate the galaxy
LF at several redshifts and rest-frame wavelenghts.  They do not
present a rest-frame 1700\AA\ LF and so we compare our KDF results to
their 1500\AA\ LF.  Both these rest-frame wavelenghts probe the part
of galaxy SEDs dominated by young, massive stars and so the systematic
biases introduced by this slight difference in wavelenghts should not
be large.

Figure~\ref{fig-LF-otherpple.fig} uses downward-pointing red triangles
to show the 1500\AA\ Gabash et al.\ (2004) FDF luminosity functions at
\zs3 and \zs4.  The agreement between the KDF and FDF is very good,
particularly when one considers that the FDF points are based on a
single moderately-sized field and so do not account for cosmic
variance and, moreover, that the FDF and KDF use very different means
of sample selection.  In particular, we note that both the FDF and the
KDF show relatively shallow faint-end slopes at both
\zs3 and \zs4 although we feel that the KDF result is much more robust
both with respect to cosmic variance (because of its larger area and
multiple independent fields) and sample selection (because of the
extensive spectroscopic verification of its selection technique).

\subsubsection{The Subaru Deep Survey}

The Subaru Deep Survey (SDS) consists of two large fields (the Subaru
Deep Field, SDF, and the Subaru XMM Deep Field, SXDF). Both fields
have been imaged with the Suprimecam large mosaic imager (600
\sqam\ per field) on Subaru and contain statistically large numbers 
of galaxies and are very unlikely to be affected by cosmic variance.
The SDS $BVRi'$ images allow the selection of \zs4 and \zs5 LBG
samples and Ouchi et al.\ (2004a) have used such selection to estimate
the LBG LFs at these redshifts.  A strong limitation of the SDS
selection of high-$z$ galaxies is that their color-color selection
technique is not calibrated spectroscopically (only a small handful of
$z$$>$1 spectroscopic redshifts is known in the SDS fields) and must
instead rely on models of galaxy colors for its definition.  In
Fig.~\ref{fig-LF-otherpple.fig} we use blue squares to plot the
\zs4 rest-frame 1700\AA\ luminosity function from the SDF field (we
omit the SXDF field which shows virtually identical results but is
$\sim$0.5 mag shallower than the SDF).  At the bright end, the SDS
\zs4 LF agrees both with the KDF luminisity function (which at these
magnitudes is dominated by the Steidel et al.\ 1999 data) and with the
FDF result. However, at the faint end, the SDS has a much steeper LF
slope than either the KDF or the FDF.  The origin of this discrepancy
is not immediately clear. One possible explanation is that at faint
magnitudes, where as we have seen in
\S~\ref{dependence-on-model-properties} the scatter in galaxy
photometry can strongly affect the estimated effective volumes used in
calculating the LF.  Ouchi et al.\ (2004a) do not discuss what galaxy
models they used to estimate their \Veff, but it is possible that
while our \Veff\ estimates at \zs4 are robust to changes in the
assumed galaxy SEDs, their effective volumes computed for their $BRi'$
color-color selection are less so.  If this is the case, then the
\Veff\ in the SDS work may well be underestimated resulting in
overcorrections to the SDS LF at the fainter magnitudes.

\section{EVOLUTION OF THE LUMINOSITY FUNCTION}\label{LFevolution}

In this section we compare our LFs at different redshifts to search
for signs of evolution.  In what follows, we concentrate on
Figs.~\ref{LF-zcomp-z4-3.fig} and \ref{LF-zcomp.fig} which compare
directly our \zs4, 3, 2.2, 1.7 LFs that were determined using the
baseline \Veff\ model (100Myr-old starburst with
\ebv=0.15).  As we will elaborate below, our \zs3 and \zs4 LFs are the 
most unambiguous and free of systematics, whereas the \zs2.2 and
especially \zs1.7 LFs are likely biased; for this reason in
Fig.~\ref{LF-zcomp-z4-3.fig} we present the \zs4 and \zs3 LFs only,
without the \zs2.2 and \zs1.7 LFs that are shown in
Fig.~\ref{LF-zcomp-z4-3.fig}.  Thus, Fig.~\ref{LF-zcomp-z4-3.fig}
shows only the most solid, bias-free results, while
Fig.~\ref{LF-zcomp.fig} should thus be regarded as presenting all the
results, including those biased by systematic effects.

Our LF is most unambiguously constrained at \zs3, where the
combination of KDF and Steidel et al.\ (1999) data covers the largest
range in luminosity, is least dependant on the details of
\Veff\ modeling (see \S~\ref{dependence-on-model-properties}), and
puts the tightest bounds on the Schechter function parameters.
Consequently, we will use the \zs3 LF as our fiducial reference and
compare the other redshift bins to it.

The top panels of Figs.~\ref{LF-zcomp.fig} and \ref{LF-zcomp-z4-3.fig}
overplot the data and the Schechter function fits, and the bottom
panels further highlights the differences between the three redshift
samples by showing the data after they have been divided by the
Schechter function fit to the \zs3 LF.  Three interesting evolutionary
effects can be seen in Figs.~\ref{LF-zcomp.fig} and
\ref{LF-zcomp-z4-3.fig}:
\begin{itemize}
\item A strong increase in the number density of {\it low-luminosity}
 LBGs from \zs4 to \zs3.
\item An accompanying apparent lack of change in the number density of 
{\it luminous} LBGs from \zs4 to \zs3.
\item A possible continuation of the increase in the number density of 
low luminosity galaxies from \zs3 to \zs2.2 and \zs1.7.
\end{itemize}
We discuss these three effects in the following subsections.  In each
subsection, we give a brief phenomenological description of the effect
and then concentrate on examining whether the evolution may be
explained as an observational effect or whether it instead reflects a
true evolution of the underlying galaxy population.  Discussion of the
possible implications for galaxy evolution of the observed LF
evolution are deferred to \S~\ref{discussion}.

\subsection{Increase in the number of low-luminosity  galaxies from \zs4 to \zs3}
\label{z4z3faintend}

The most striking effect in Fig.~\ref{LF-zcomp-z4-3.fig} is the
increase in the number density of low-luminosity galaxies from \zs4 to
\zs3.  For our adopted fiducial \ebv=0.15 model, the number of
\zs4 galaxies with $-$21$<$\Muv$\lesssim$$-$18 (i.e, within
\Mstar$\lesssim$\Muv$\lesssim$\Mstar+2) is only 0.44$\pm$0.05
of the number of such galaxies at \zs3.  In other words, the number of
low luminosity galaxies increases by a factor of 2.3 from \zs4 to \zs3
--- a result that is statistically significant at the $\sim$11$\sigma$
level.

Is this increase a reflection of true galaxy evolution, or is it
simply an artefact of some observational bias?

\subsubsection{Can cosmic variance be responsible?}  

The effects of large scale structure can be a problem in small-area
surveys such as the HDFs or those that consist of a single pointing,
such as the FORS Deep Field.  The KDF, however, consists of three
widely-separated patches on the sky that probe statistically
independent parts of the Universe.  As was discussed in
\S\S~\ref{LFcalculation} and
\ref{theLF}, field-to-field fluctuations are relatively small and 
the errorbars in our LF measurement include a bootstrap estimate of
field-to-field variance of our survey.  Given that the increase in the
number of low-luminosity galaxies from \zs4 to \zs3 is far outside
these errorbars, we conclude that cosmic variance is unlikely to be
responsible for the observed evolution of the faint end of the LF.

\subsubsection{Is our modeling of \Veff\ responsible?}  

As we discussed in \S~\ref{dependence-on-model-properties}, the
computation of the LF is dependent on the details of the \Veff\
modeling of the survey.  However, as we have shown in
\S~\ref{dependence-on-model-properties}, at \zs4 and \zs3 the 
change in the LF is remarkably small under a wide range of reasonable
assumptions.  As is illustrated in the top panel of
Fig.~\ref{LF-4panel-modelcomp.fig}, varying the \Veff\ model
assumptions cannot bring the faint ends of the \zs3 and \zs4 LF into
agreement.  And as Fig.~\ref{fixedphi-vs-model.fig} shows further, it
appears to be impossible to bring the number density of $\sim$\Mstar+1
galaxies at \zs3 and \zs4 into agreement by varying \Veff\ model
parameters: even in the most extreme \Ebv=0 case, where the number
densities at \zs4 and 3 are closest to each other, there remains a
very significant deficit of faint galaxies at \zs4.  We therefore
conclude that the evolution of the faint end seen in our data is
unlikely to be an artefact of the assumptions that underlie our
calculation of \Veff.

\subsubsection{Can differential sample selection be responsible?}  

\begin{figure}
\plotone{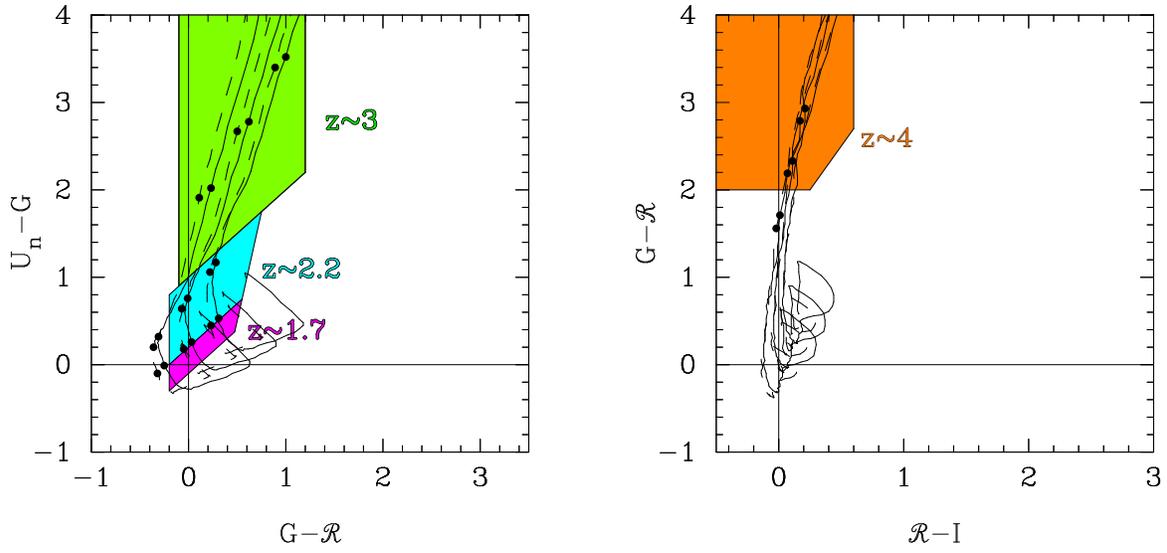}
\caption{\label{colorcolor.fig}
Colors of model galaxies in the \UGR\ (left panel) and \GRI\ (right
panel).  The filled regions represent the color-color selection
criteria used to select our samples of high-$z$ galaxies (see
\kdfdata\ and Steidel et al.\ 1999; 2003; 2004). The tracks show model
colors of 100~Myr-old (solid lines) and 10~Myr-old (broken lines)
starbursts for three values of reddening each, \ebv=0, 0.15, and 0.3.
Reddening generally increases from lower left to upper right.  The
points mark the location of $z$=1.7, 2.2, 3, and 4 on each of the
tracks.  }
\end{figure}

\begin{figure}
\includegraphics[width=8cm]{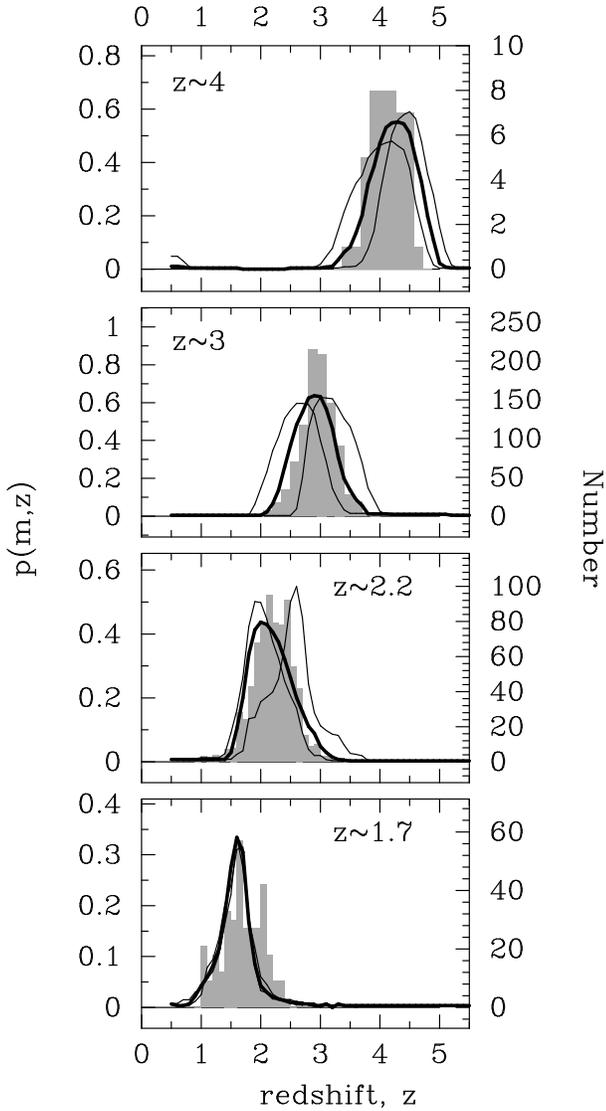}
\caption{\label{zdistr.fig}
Comparison of our \pmz\ models with the observed redshift
distributions of UV-selected high-$z$ galaxies.  The shaded histograms
show the spectroscopic redshift distributions of Steidel et al.\
(1999, 2003, 2004), while the solid curves show our \pmz\ for \ebv=0,
0.15, 0.3.  The \ebv=0.15 fiducial model is marked with a heavier line
and generally, in each panel the peak of the \pmz\ moves to lower
redshifts with increasing \ebv. These \pmz\ models are for \R=26.5 ---
a magnitude at which typical photometric errors in the KDF correspond
to those in the observed spectrocopic samples whose photometry is
shallower than ours.  Here, results for our 09A field are shown but
the other four fields give similar curves that show good agreement
between our fiducial model model and the observed redshift
distribution.}
\end{figure}

\begin{figure}
\includegraphics[width=15cm]{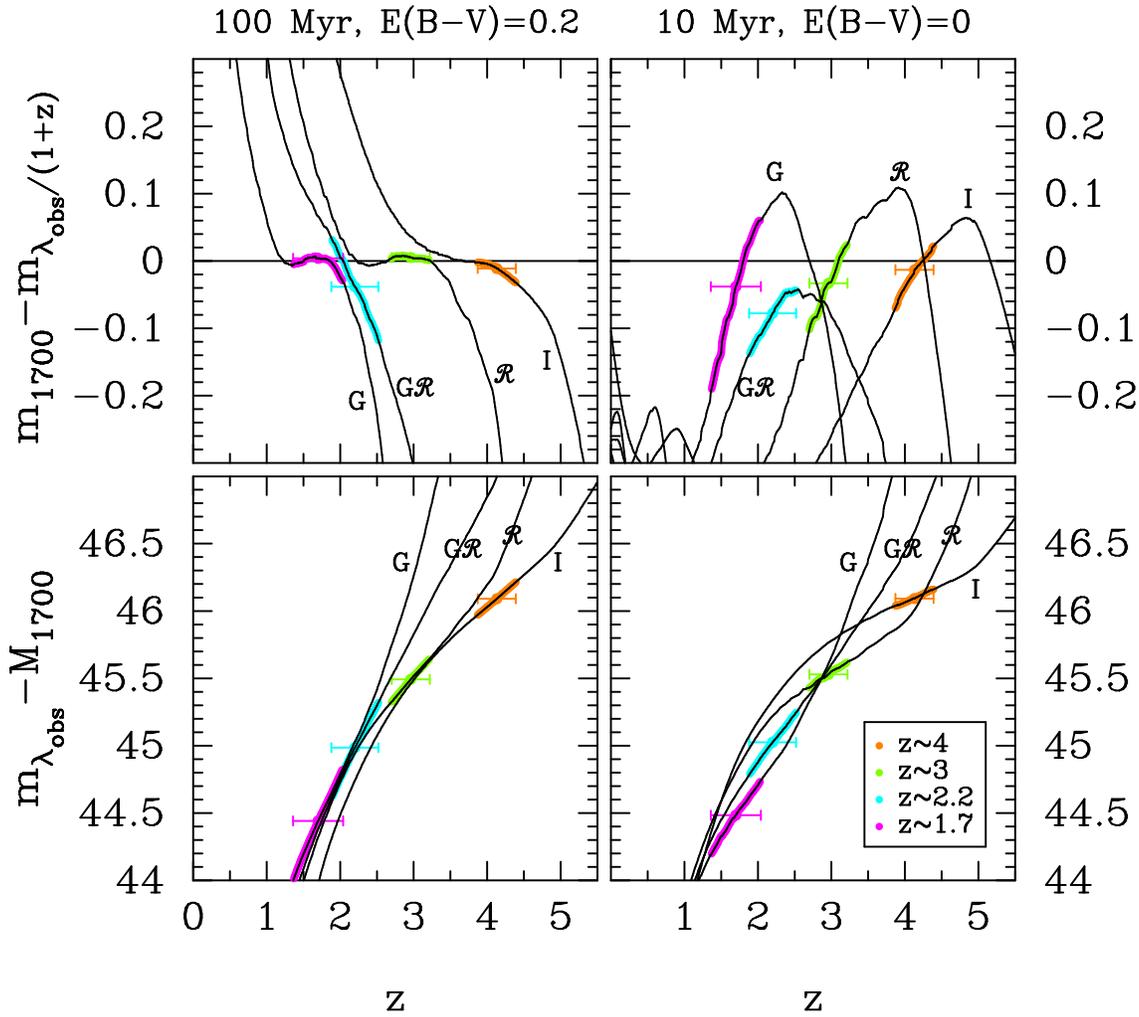}
\caption{The $k$-correction\label{kcorr.fig} 
color term ({\it top panels}) and the distance moduli ({\it bottom
panels}) for two representative galaxy SEDs.  Locations of the \zs4,
\zs3, \zs2.2, and \zs1.7 sample bin centres are marked. The {\it top
panels} show the color terms that are needed to transform observed \G,
\GRave, \R, and \I\ magnitudes to rest-frame 1700\AA.  The top left
panel shows the color term for a 100 Myr-old starburst with moderate
dust attenuation, while the right panel is for an unobscured 10 Myr
starburst.  The errorbars show the FWHM redshift ranges ($\delta
z$$\sim$0.3; see \kdfdata\ for details) spanned by the color-color
selected samples, and the thick colored bands highlight the color term
values corresponding to these redshift ranges.  These plots illustrate
that for the right choice of observed bandpass --- namely \I\ for \zs4
LBGs, \R\ for \zs3 ones, \GRave\ for \zs2.2, and \G\ for \zs1.7 ---
the $k$-correction color is $\sim$0.  The {\it bottom panels} show the
corresponding offsets between observed and absolute magnitudes (as
defined in Eq.~\ref{m2M.eq} --- i.e., including the $k$-correction
color term), and illustrate that the redshift uncertainty for our
photometrically-selected objects translates into only a small DM
uncertainty, and hence into only a small uncertainty in the derived
absolute UV magnitude of the object, $M_{1700}$.}
\end{figure}

\begin{figure}
\includegraphics[width=7cm]{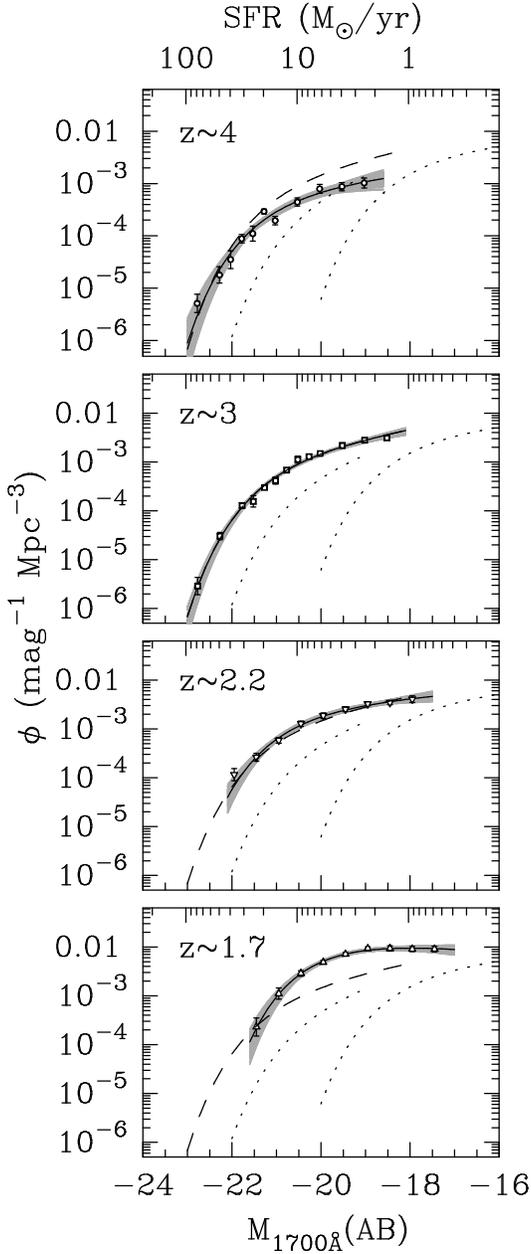}
\caption{\label{lf4pan.fig}
The rest UV luminosity functions at different redshifts.  The solid
black curves show the best-fitting Schechter functions, while the
shaded bands show the corresponding 68.3\% confidence region.  As is
described in the text, the errorbars include both the $\sqrt{N}$
statistics and a bootstrap estimate of field-to-field variations.  The
dashed fiducial curve simply reproduces the \zs3 curve.  The dotted
curves show the GALEX rest-frame 1500\AA\ LFs for comparison; the
rightmost, fainter one is for \zs0 and the leftmost, brighter one is
for \zs1.  As we discuss in the text, we regard the \zs4 and 3 LFs
shown here to be highly trustworthy, the \zs2.2 to represent a firm
lower limit on the galaxy number density, but the \zs1.7 to be
questionable due to systematic biases.  }
\end{figure}

\begin{figure}
\includegraphics[width=8cm]{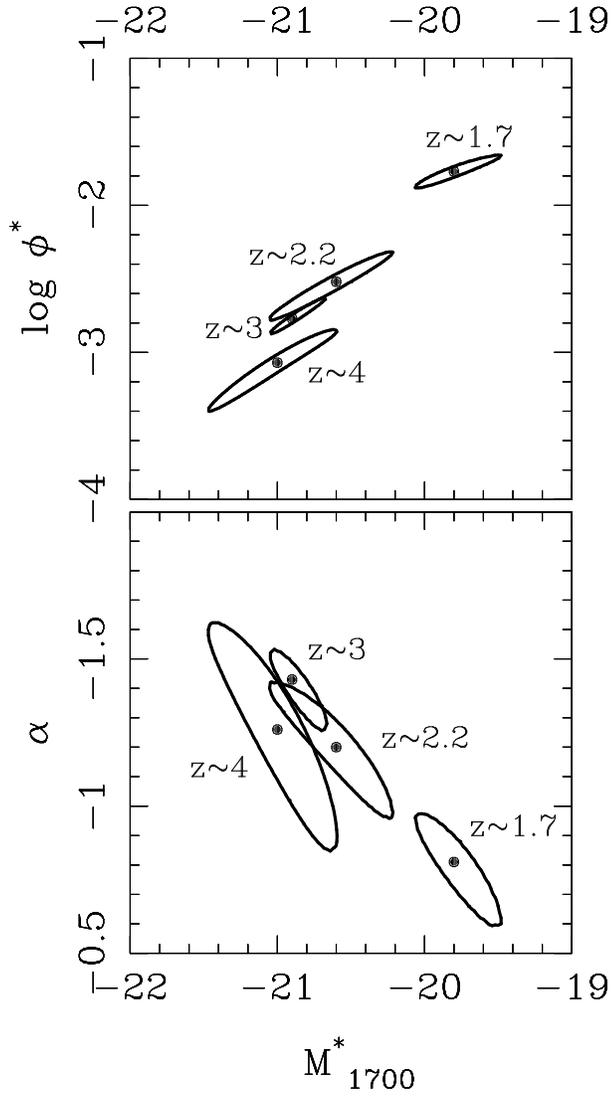}
\caption{\label{schechter.fig}
The Schechter luminosity function parameters and their 1$\sigma$
confidence regions.  }
\end{figure}

\begin{figure}
\includegraphics[width=10cm]{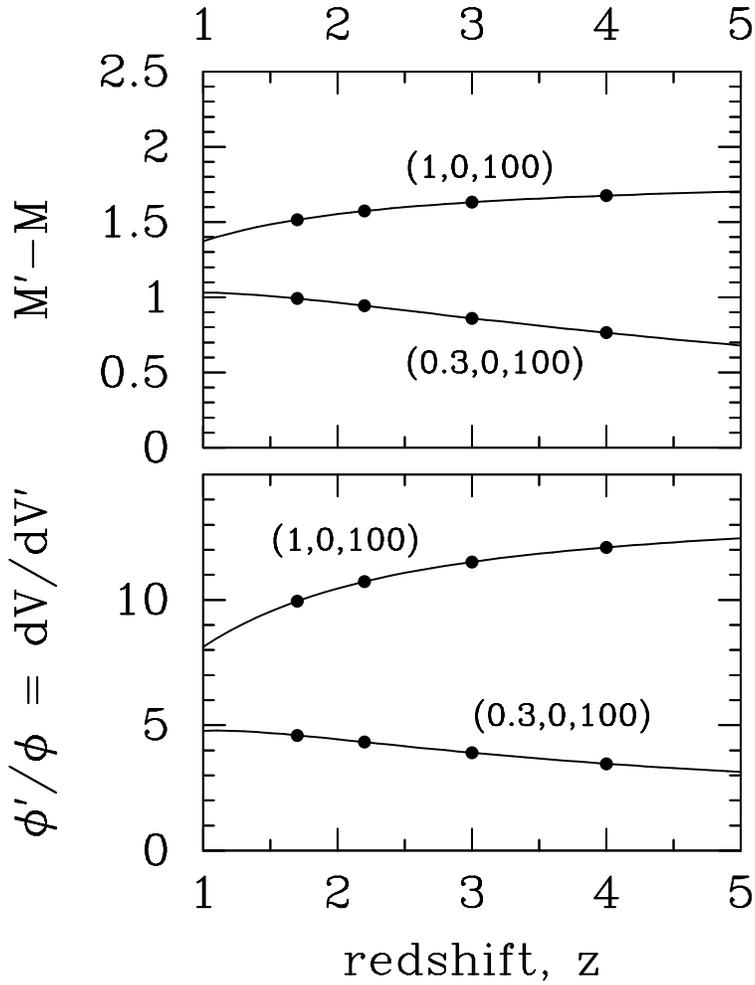}
\caption{\label{comparecosmo.fig}
The effect of changing cosmology on absolute magnitudes (top panel)
and number densities (bottom panel) as a function of redshift. The
quantities plotted show the change incurred in transforming from the
cosmology adopted in this paper --- \OmOlHo=(0.3,0.7,70) --- to the
two alternative cosmologies labeled in the plots.  The values of $M$
and $\phi$ in the alternate cosmologies are denoted with primed
quantities while in our cosmology they are unprimed. While the change
in $M$ or $\phi$ at any given redshift can be quite large, the {\it
relative} change from redshift to redshift is small, ensuring that ---
for reasonable cosmologies such as those considered here --- any
evolutionary trends seen in the LF are not subject to the assumed
cosmology.}
\end{figure}

\begin{figure}
\includegraphics[width=8cm]{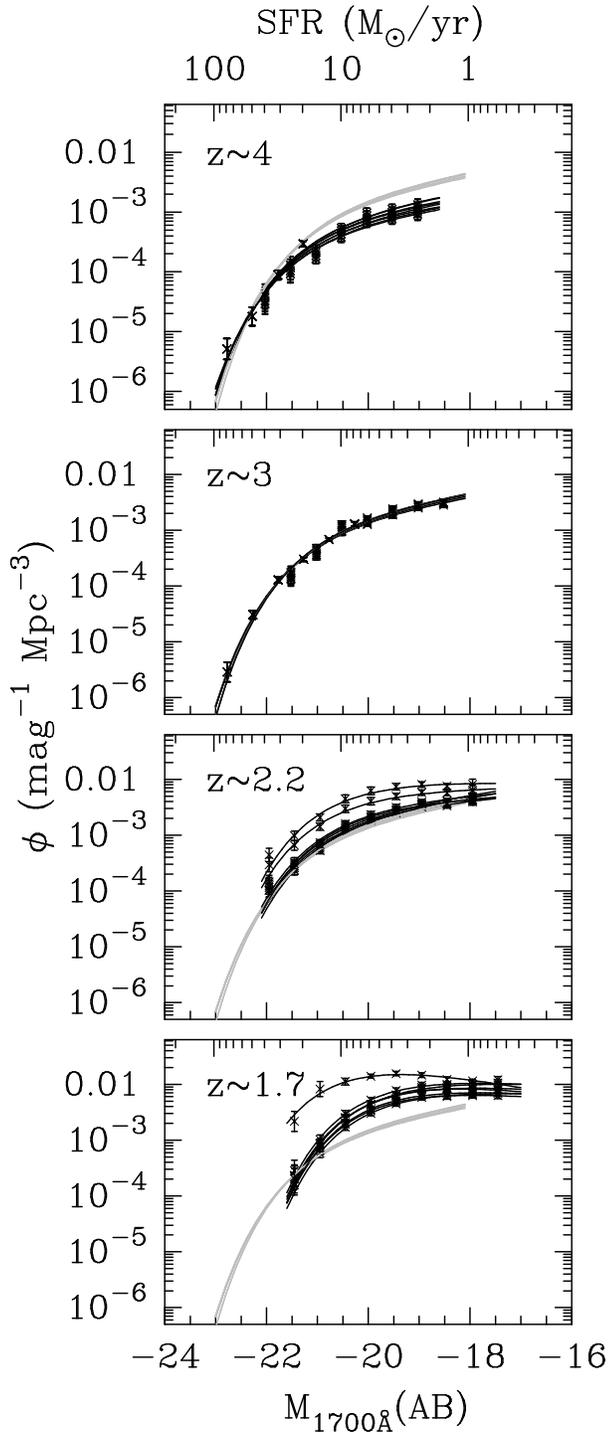}
\caption{\label{LF-4panel-modelcomp.fig}
Dependence of the LF on the \Veff\ model.  Variation in the derived LF
for the range of considered \Veff\ models.  The points and black
curves show the results of recomputing the LFs using the seven
different models of \Veff\ that result from seven different \Ebv
values. The gray curves are the \zs3 results replotted in the other
redshift panels.  See text for more details.}\end{figure}

\begin{figure}
\plotone{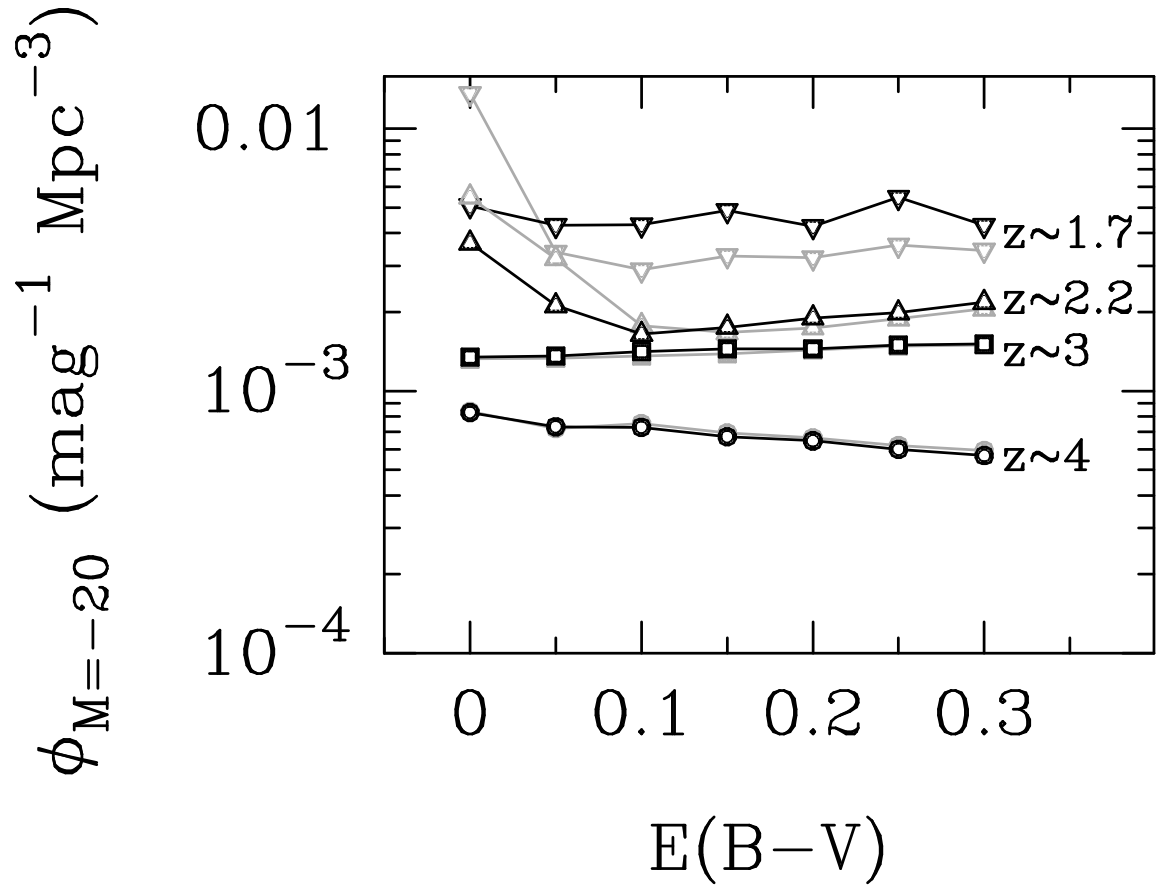}
\caption{\label{fixedphi-vs-model.fig}
Dependence of the galaxy number density at \Muv=$-$20 on the \Veff\
model.  The black symbols show 100~Myr-old starburst models, and the
gray ones are for 10~Myr-old models.  }\end{figure}

\begin{figure}
\includegraphics[width=10cm]{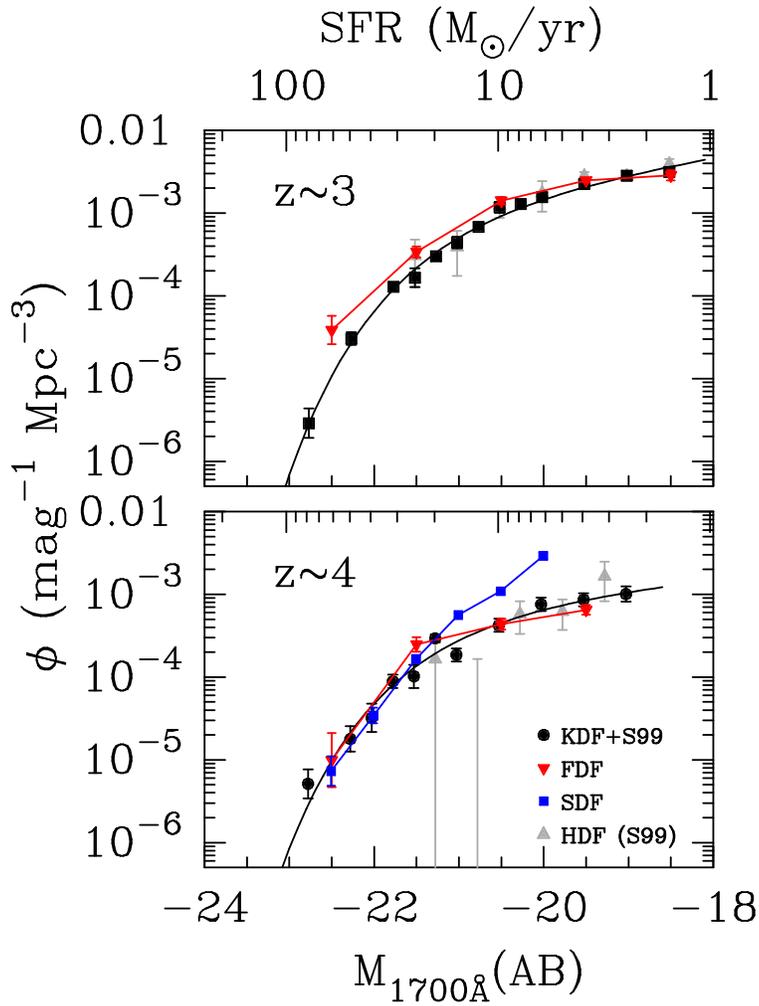}
\caption{\label{fig-LF-otherpple.fig}
Comparisons with other surveys.  Rest-frame UV luminosity functions
measured in the FORS Deep Field (FDF; Gabasch et al.\ 2004) at \zs3
and \zs4 and in the Subaru Deep Field (SDF; Ouchi et al.\ 2004a) at
\zs4 are shown.  Also shown are the results of the Steidel et al.\ 
(1999) analysis of the HDF.  The Lyman Break Galaxy LF is shown as
filled circles with errorbars that include an estimate of the
field-to-field variance determined via bootstrap resampling.  The FDF
measurement is based on a single small field and does not include an
estimate of cosmic variance while the SDF measurement is based on a
single very large field and so is unlikely to be strongly affected by
cosmic variance.  The LBG luminosity function is in very good
agreement with that found in the smaller-area FDF at both \zs3 and
\zs4.  However, while the KDF and the FDF agree with each other, they
both disagree with the SDF measurement at \zs4.  }
\end{figure}

\begin{figure}
\plotone{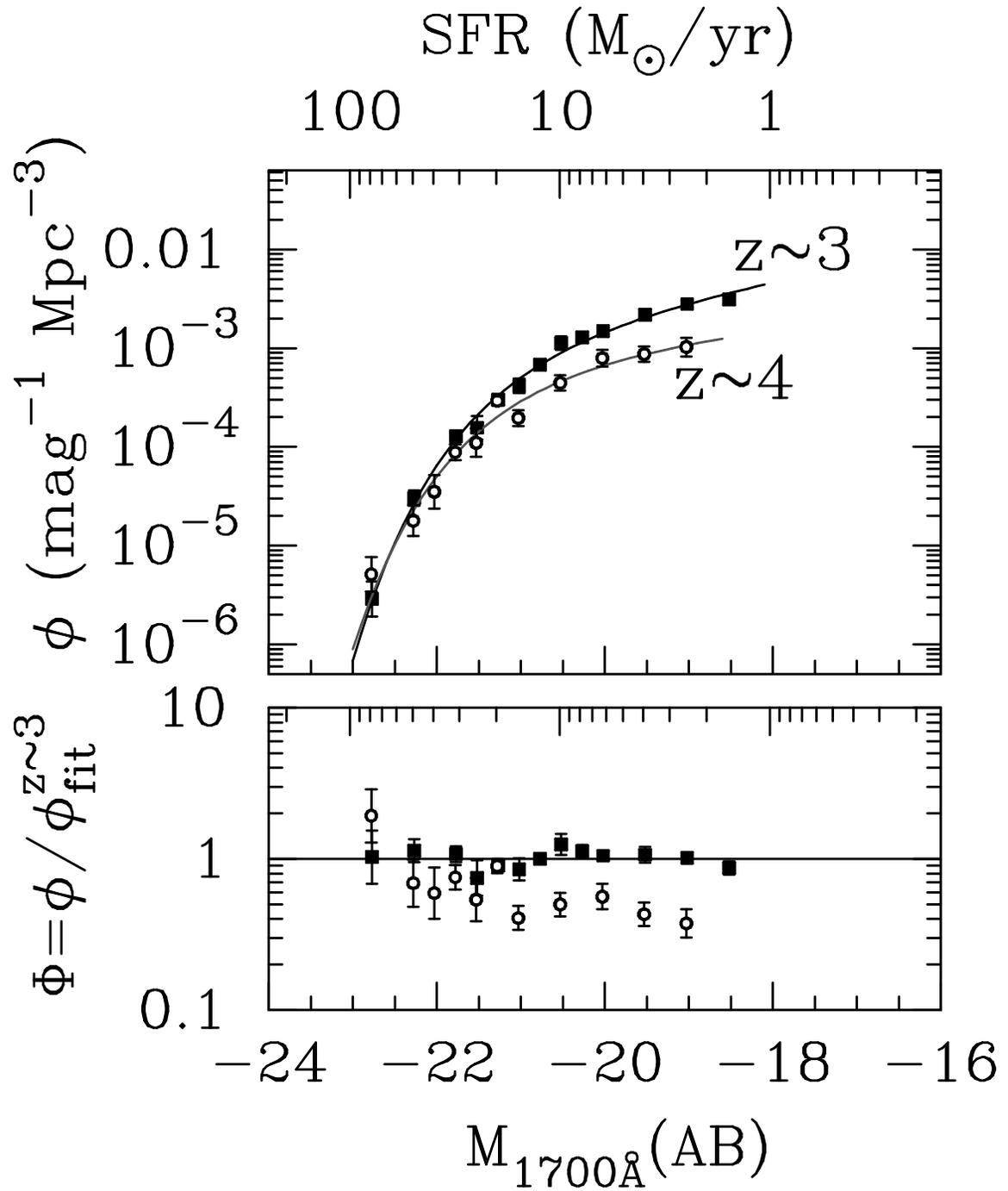}
\caption{\label{LF-zcomp-z4-3.fig}
Comparison of the \zs4 and \zs3 luminosity functions.  The top panel
shows the \zs4, LFs together, with the \zs3 Schechter fit marked with
a heavier line.  The bottom panel shows the quantity $\Phi$, which
measures the fractional deviation of the data (at \zs4 or 3) from the
\zs3 Schechter fit: perfect agreement between data and the
\zs3 fit would put the points on the horizontal $\Phi$=1 line. }
\end{figure}

\begin{figure}
\plotone{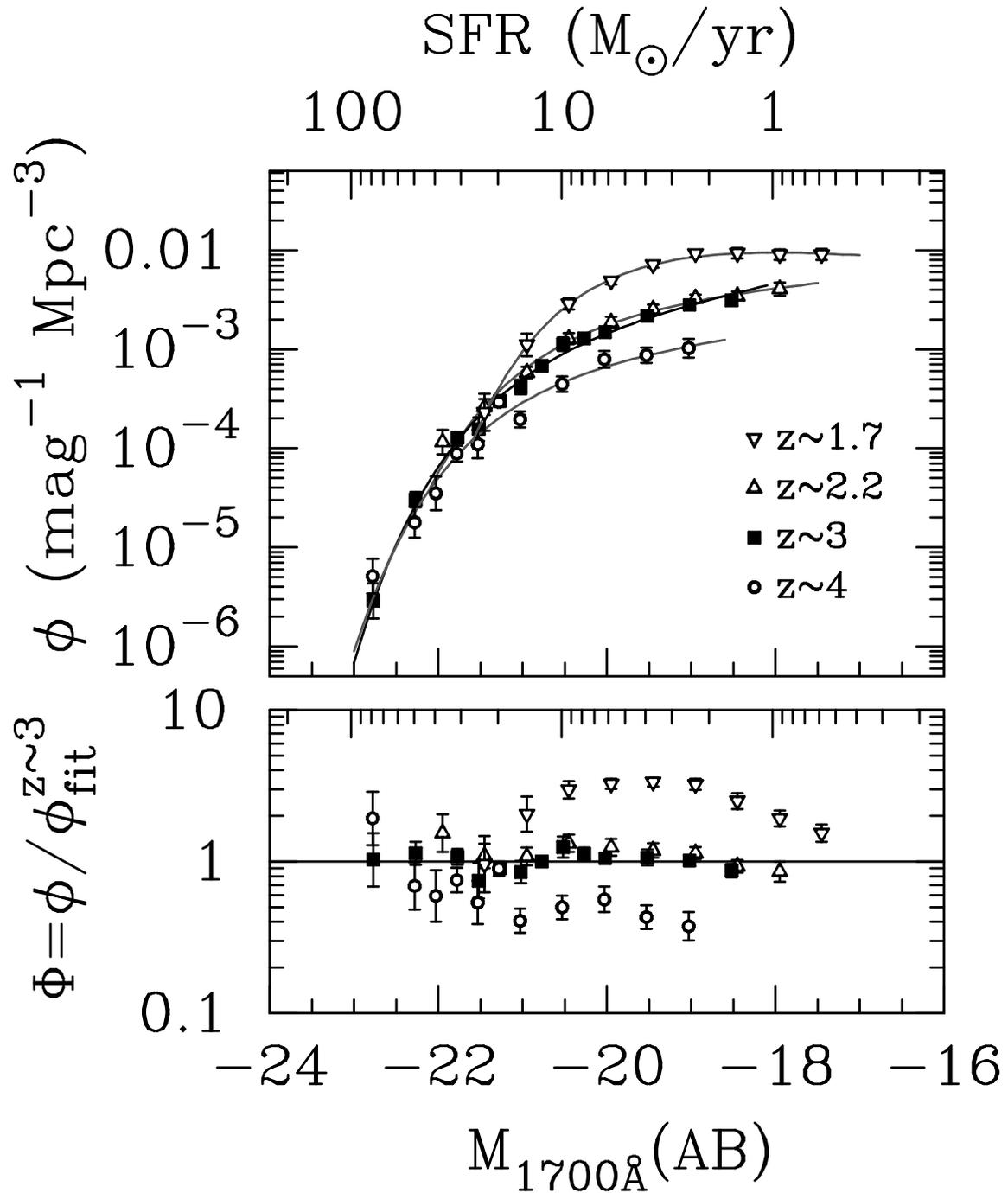}
\caption{\label{LF-zcomp.fig}
Comparison of the \zs4, 3, 2.2, and 1.7 luminosity functions.  As in
Fig.~\ref{LF-zcomp.fig} but adding the \zs2.2 and 1.7 LFs.  As is
discussed in the text, the \zs4 and \zs3 LFs are robust to the details
of the analysis procedure, whereas the \zs2.2 and \zs1.7 are less so
and are subject to potentially strong systematic uncertainties.  }
\end{figure}

The color selection criteria used to select galaxies at \zs3 and
\zs4 are designed to select galaxies with similar underlying SEDs.  
However, the color-color selection regions are {\it not} identical in
relation to intrinsic galaxy colors (see Fig.~\ref{colorcolor.fig})
and so we may question whether the deficit of \subLstar\ galaxies at
\zs4 results simply because at \zs4 we are missing galaxies whose SEDs
are such that they would have been included in the \zs3 selection
criteria.

If there are significant numbers of such galaxies missing from our
sample at \zs4 then we should be able to include them by expanding the
\zs4 color-color selection region.  However, the deficit at the faint 
end is so large --- a factor of 2.3 --- that no reasonable adjustment
to the \zs4 color selection criteria can remedy it.  As
Fig.~\ref{colorcolor.fig} shows, the galaxy color model tracks allow
the possibility of modifying the \zs4 color-color selection region by
a few tenths of a magnitude.  However, we have checked that an
increase of even as much as 0.5 mag in both \GR\ and \RI\ would not be
sufficient to bring enough galaxies into the sample to make up the
factor-of-2.3 deficit (see also Fig.~4 in \kdfdata).  Moreover, even
if such a large change to the color selection criteria {\it were}
permissible, it would result in an automatic increase of the effective
volume \Veff\ that would largely counteract any gain from the increase
in galaxy numbers.

A further, secondary argument against differential sample selection
relies on the fact that the evolution from \zs4 to \zs3 appears to be
differential with luminosity (see \S~\ref{differentialevolution}).
Since at a given redshift bright and faint LBGs are selected in {\it
the same} way, a sample selection bias should result in a similar
deficit of bright galaxies as of faint ones at \zs4.  The fact that no
strong deficit is seen at the bright end of the \zs4 LF further
strengthens the case that the evolution of the faint end is not an
artefact of sample selection but is due to real differences between
luminous and low-luminosity galaxies.

Overall, we believe that the deficit of faint galaxies at \zs4 is too
large to be accounted for by differences in sample selection and is
most likely a real effect that reflects an underlying evolutionary
change in the population of sub-\Lstar galaxies over the $\sim$600 Myr
from \zs4 to \zs3.  We conclude that the observed evolution of the
faint end of the LF from \zs4 to \zs3 is likely real.

\subsection{Luminosity-dependent evolution from  \zs4 to \zs3}
\label{differentialevolution}

Figure~\ref{LF-zcomp-z4-3.fig} suggests that while there is a deficit
of \zs4 \subLstar\ galaxies, the number of luminous galaxies remains
virtually unchanged from \zs4 to \zs3.  If real, this differential,
luminosity-dependent evolution of the LF hints at important
differences in how galaxies of different luminosity evolve at high
redshift.  We examine the potential implications of this differential
evolution in \S~\ref{discussion}, but first, here, we ask if this
differential effect is in fact real.

\subsubsection{Is the effect statistically significant?}  

While there can be little doubt of the deficit of {\it faint} galaxies
at \zs4, the situation at the bright end is less clear because of the
larger uncertainties in the individual data points.  In view of this
problem, it is tempting to rely on the Schechter functions to compare
the bright-end LFs, and indeed the \zs3 and \zs4 Schechter function
fits are in excellent agreement at the bright end.  However, such a
comparison can be misleading because the Schechter function fits
incorporate data at {\it all} luminosities and so at the bright end
the fits may well be biased by the weight of the faint-end data where
the statistics are so much better.  It would be more robust to compare
the \zs3 and \zs4 number densities directly, but this is not
straightforward because the data at \zs4 and \zs3 are sampled
differently and cover somewhat different magnitude ranges in the two
redshift bins.

To overcome these difficulties, we instead compare how the data at
\zs3 and \zs4 deviate from our \zs3 Schechter function fit.  We
proceed as follows.  First, we calculate the quantity \PhiM, which is
the ratio between the data and the \zs3 Schechter function fit,
\begin{equation}
\Phi(M) = \phi_{data}(M) / \phi^{z\sim3}_{fit}(M).
\end{equation}
The bottom panels of Figs.~\ref{LF-zcomp.fig} and
\ref{LF-zcomp-z4-3.fig} show \PhiM\ for our four redshift samples.
Note that \PhiM\ is computed for each of the four redshift samples,
but in each case the data are divided by the same, \zs3, Schechter
function.  Consequently, \PhiM\ is close to 1 at
\zs3 (reflecting the fact that the \zs3 Schechter function is a good
representation of the \zs3 data) but deviates from 1 for the other
redshift bins, and in particular for the \zs4 sample.  We then compute
the {\it average} \PhiM, namely
\Phibar, for galaxies brighter and fainter than \Muv=$-$21.0 
(i.e., $\sim$\Mstar) at both redshifts of interest here, \zs3 and
\zs4.  Finally, the {\it ratio} of the \Phibar\ at the {\it two
redshifts} then tells us the amount of number density evolution that
the given population undergoes.  By comparing the {\it ratios} of the
\Phibar, we effectively cancel out the dependence of our comparison on
the \zs3 Schechter function fit and are comparing the data at \zs3 and
\zs4 directly.

\begin{figure}
\plotone{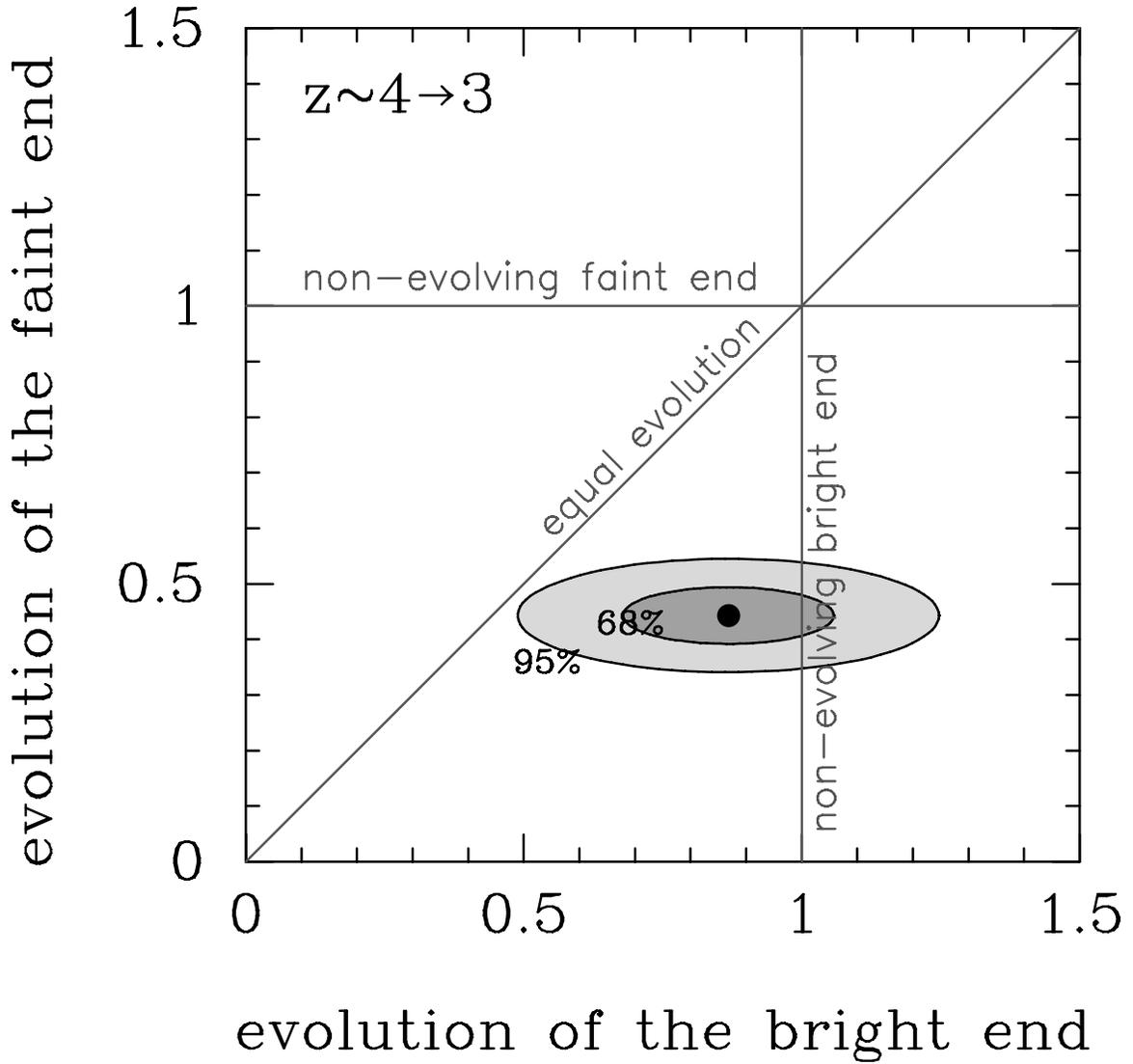}
\caption{\label{faint-vs-bright-LFresiduals.fig}
The luminosity-dependent evolution of the number density of galaxies.
The horizontal axis shows the amount of change in the density of
luminous galaxies (\Muv$<$$-$21) parametrized as the LF as the ratio
between the average normalized densities $\bar{\Phi}$ at
\zs4 and \zs3. The vertical axis shows the same quantity 
but for low-luminosity galaxies, \Muv$>$$-$21.  The straight lines
represent three fiducial cases: no faint-end evolution (horizontal
line), no bright-end evolution (vertical line) and equal evolution at
the bright and faint ends (diagonal line).  The point and ellipses
represent the amount of evolution from \zs4 to \zs3 and the associated
68\% and 95\% confidence regions.  There is substantial evolution of
the low-luminosity population from \zs4 to \zs3 that is statistically
significant at the 98.7\% level.  }\end{figure}

Figure~\ref{faint-vs-bright-LFresiduals.fig} shows these
\Phibar\ ratios for the bright (horizontal axis) and faint (vertical
axis) ends of the luminosity function.  The quantities in
Fig.~\ref{faint-vs-bright-LFresiduals.fig} are always shown as
evolution with respect to the \zs3 case (i.e., we plot
\Phibar$_z$/\Phibar$_{z\sim3}$).  Three evolutionary scenarios
are marked for reference: locations on the vertical straight line
indicate no number density change in the bright end of the LF,
locations on the horizontal solid line indicate no evolution in the
faint end, and locations on the diagonal line indicate equal number
density evolution at the bright and faint end.  The intersection of
the three lines at (1,~1) marks the case of a non-evolving LF.

As Fig.~\ref{faint-vs-bright-LFresiduals.fig} shows, there is
substantial change in the number density ratio of sub-\Lstar\ galaxies
from \zs4 to \zs3: \Phibar(\zs4)/\Phibar(\zs3)=0.44$\pm$0.05 for {\it
faint} galaxies, indicating that there is a 2.3-fold increase in the
number density of faint galaxies that is statistically significant at
the $\sim$11$\sigma$ level (this is the faint-end evolution we
discussed in \S~\ref{z4z3faintend}).  At the same time, however,
Fig.~\ref{faint-vs-bright-LFresiduals.fig} also shows that the number
density of {\it luminous} galaxies at \zs4 is virtually unchanged with
\Phibar(\zs4)/\Phibar(\zs3)=0.87$\pm$0.19.

To properly test whether the evolution of the LF is differential with
luminosity, we must of course consider the {\it joint} uncertainty for
the bright- and faint-end cases.  The joint 1$\sigma$ and 2$\sigma$
uncertainties are illustrated by the error ellipses in
Fig.~\ref{faint-vs-bright-LFresiduals.fig}, and in this context, the
distance of the \zs4$\rightarrow$3 point from the diagonal ``equal
evolution'' line indicates the amount of differential evolution.  We
find that 98.5\% random realizations of the data in
Fig.~\ref{faint-vs-bright-LFresiduals.fig} are above the diagonal
``equal evolution'' line and so inconsistent with the differential
evolution scenario.  This strongly suggests that the high-$z$ galaxy
population is undergoing differential, luminosity-dependent evolution.
Stronger confirmation of this assertion will require improved
constraints on the bright end of the LF; this confirmation will
require LBG surveys with areas of several square degrees --- i.e.\ an
order of magnitude greater than used in the work of Steidel et al.\
(1999) that provides the bulk of the statistics at the bright end of
the LF here.

\subsubsection{Can differential sample selection be responsible?}  

The bright and faint galaxies in a given redshift bin are selected
using identical selection criteria.  Moreover, given that the
bright-end galaxies come primarily from the Steidel et al.\ (1999)
work and the faint-end ones exclusively from the significantly deeper
KDF, they suffer from similar amounts of photometric scattering out of
the color-color selection regions (at any rate, this scatter is
accounted for through the \Veff\ approach).  Consequently, as long as
the bright and faint samples within that bin consist of the same mix
of galaxy SEDs, any systematic effect should be reflected in the same
way at both the bright and faint ends.

At the bright end, the LF is constrained mainly by the Steidel et al.\
(1999) results, while the faint end is dominated by the KDF data.  It
is possible that despite our great care some subtle, unknown selection
effect remains between the two samples.  However, we consider this
possibility to be extremely unlikely in view of the virtually
identical observational approaches and sample selection and analysis
techniques that were applied to the two datasets.

Finally, we cannot rule out the possibility that the faint and bright
galaxies at a given redshift represent populations that differ not
just in luminosity but in some other property such as, for example,
the amount of interstellar dust.  However, if such true, intrinsic
luminosity-dependent differences do exist, they do not alter our
conclusion that the LF evolution is differential with luminosity, but
merely shift the focus to a more specific, but still real,
evolutionary effect.  We conclude that while the different composition
of the samples {\it with luminosity within a given redshift bin} may
explain the differential, luminosity-dependent evolution we see from
\zs4 to \zs3, such differences in intrinsic galaxy properties would
only underscore the point that the luminous and faint UV-selected
galaxy sub-populations {\it are} different and not simply scaled
analogs of each other.

\subsection{Evolution from \zs3 to \zs2.2}

At \zs2.2 (and \zs1.7) we do not have sufficient statistics in the KDF
data to say much about the LF for galaxies brighter than \Lstar, nor
--- unlike at \zs3 and \zs4 --- are there published results from the
Steidel group that we could use to extend our luminosity range.
Consequently, we focus on the evolution of the faint end of the LF
only.

Figure~\ref{LF-zcomp.fig} shows evolution of the LF from \zs3 to
\zs2.2 (see also Figs.~\ref{lf4pan.fig} and \ref{LF-4panel-modelcomp.fig}).  
The LF appears to undergo only a small amount of evolution from \zs3
to \zs2.2 for our fiducial \Veff\ model (100~Myr-old starburst with
\ebv=0.15):  there is only a factor of 1.1$\pm$0.1 increase in the number 
density of sub-\Lstar\ with cosmic time.  However, as
Fig.~\ref{fixedphi-vs-model.fig} illustrates, our chosen fiducial
\Veff\ model happens to give nearly the minimal evolution in the LF from 
\zs3 to \zs2.2 and applying \zs2.2 \Veff\ models with either larger or smaller
extinction values results in stronger LF evolution.  For example,
using the \ebv=0.3 model for both \zs3 and \zs2.2 gives a factor of
$\sim$1.2 increase in the number density of sub-\Lstar\ galaxies and a
yet stronger evolution (factor of 2.3 number density increase) results
with the \ebv=0 model.  Likewise, adopting a \Veff\ model which
assumes a galaxy population with a {\it range} of \ebv\ values gives
evolution that is stronger than our fiducial case: using a flat number
distribution over \ebv=0--0.3 results in an increase of 1.24$\pm$0.08
in the number density of galaxies fainter than \Muv=$-$21.0.

The above examples illustrate that the measurement of the \zs2.2
luminosity function has a systematic uncertainty that depends on our
assumptions about the properties of the UV-bright galaxy population at
these redshifts.  Whereas the \zs4 and, especially, \zs3 LFs are only
weakly dependent on the assumed \Veff\ model, the \zs2.2 LF shows a
much stronger dependence.  A direct result of this dependence is that
we cannot unambiguously determine the \zs2.2 LF and the amount of LF
evolution from \zs3 to \zs2.2.

Despite these limitations, we can still put some useful constraints on
the evolution of the LF over the 800~Myr from \zs3 to \zs2.2.  As
Fig.~\ref{LF-4panel-modelcomp.fig} shows (see also
Fig.~\ref{fixedphi-vs-model.fig}), there are {\it at least} as many
galaxies at a given UV luminosity at \zs2.2 as there are at \zs3.
Depending on the adopted \Veff\ models, the number density can be
close to identical at the two redshifts, but potentially can be up to
a factor of $\sim$2 higher at \zs2.2 than at \zs3. Note that this
conclusion could potentially be further affected by differential
sample selection between \zs3 and \zs2.2, although it is unlikely that
sample selection differences will drastically modify the results given
that the \zs3 and \zs2.2 color-color selection criteria select
galaxies with similar intrinsic SEDs (Fig.~\ref{colorcolor.fig}) and
are likely to capture the bulk of UV-bright galaxies at these
redshifts, as can be seen in Fig.~5 of \kdfdata.

In summary, we can conclude that there are probably {\it at least} as
many --- and possibly more --- sub-\Lstar\ UV-bright galaxies at
\zs2.2 as there are at \zs3.  There is {\it no} evidence for a decline in
galaxy numbers with cosmic time.

\subsection{Towards \zs1.7?}\label{z17evolution}

Figure~\ref{LF-zcomp.fig} shows apparent strong evolution of the LF
from higher redshifts down to \zs1.7.  However, we feel that the LF
measurement at \zs1.7 is not reliable given systematic uncertainties
in our LF estimation at this redshift.  The main systematic problem
stems from the lack of robustness in the modeling of \Veff\ for the
\zs1.7 sample.  While we have high confidence that our \Veff\
modeling is robust at \zs4 and \zs3, and somewhat less accurate but
still partially reliable at \zs2.2, we have far less confidence in its
reliability at \zs1.7, as explained below.

While our \Veff\ modeling for the higher redshift bins accurately
reproduces the observed redshift distribution of galaxies at \zs4, 3,
and 2.2, it fails to do so at \zs1.7, as can be seen in
Fig.~\ref{zdistr.fig}.  Another manifestation of this problem can be
seen in Fig.~\ref{colorcolor.fig}, where the model colors of $z$=1.7
galaxies ($z$=1.7 is the mean redshift of the observed spectroscopic
sample) are too red in \UG\ compared to the color-color selection
region defined by Steidel et al.\ 2004.  The upshot of these
mismatches is that we do not have confidence that our modeling of
\Veff\ accurately reflects the true volume from which the color-color
selected galaxies in the \zs1.7 sample are drawn.  An inaccurate
calculation of the survey volume has the potential to strongly affect
the computed luminosity function. If, for example, our \Veff\ modeling
underestimates the volume of the survey, then this will translate into
an overestimate of the number density of galaxies.  This may well be
happening and would explain the high density normalization of the
\zs1.7 LF seen in Fig.~\ref{LF-zcomp.fig}.

There are several possible underlying reasons why our \Veff\ modeling
at \zs1.7 may be inaccurate.  We do not at present know what \ebv\
value or star formation history are appropriate for \zs1.7 galaxies.
At \zs3, good estimates of these quantities are known from
observations (see
\S~\ref{modelcolors}), and it's not unreasonable that similar
\ebv\ and age values are also applicable at \zs4 and \zs2.2 given 
that these redshift bins are only 0.6~Gyr and 0.8~Gyr away from \zs3.
By \zs1.7, however, 1.8~Gyr after \zs3, the galaxy population may have
significantly different properties than it does at \zs3 as galaxies
evolve towards the more prosaic, less star-burst dominated and less
dust-obscured galaxy population observed by \zs1.  Assuming the \zs3
reddening and starburst age values for the \zs1.7 population may well
strongly skew our \Veff\ estimate and be reflected in the mismatches
seen at \zs1.7 in Figs.~\ref{colorcolor.fig} and \ref{zdistr.fig}.

In addition to systematic problems with the \Veff\ calculation, a
second potential systematic effect may be affecting our LF measurement
at \zs1.7: it is possible that our \zs1.7 color-color selected sample
may be significantly contaminated by low-$z$ galaxies.  While the
spectroscopically-determined contamination rate in the
\R$\sim$24-25.5 \zs1.7 sample of Steidel et al.\ (2004) is less than
5\%, even small systematic offsets in the color-scale calibration
between our and their photometric systems can result in a drastically
increased contamination fraction.  We stress here that our modeling
of \Veff\ accounts for the scatter of high-$z$ galaxies between the
different high-$z$ color-color selection regions, but does not account
for the scatter of low-redshift ($z$$\lesssim$1) galaxies into the
high-$z$ selection boxes.  At \zs4, 3, and 2.2, the scatter from low
redshift is unlikely to be an issue given that these selection regions
are far-removed from the $z$$<$1 loci in color-color space.  In
contrast, the \zs1.7 selection region is in close proximity to the
region of space dominated by $z$$\lesssim$1 galaxies and it is
possible that the low-$z$ interloper contamination rate may be quite
high. A high foreground contamination fraction would result directly
in an overestimate of the number density of \zs1.7 galaxies and an
inaccurate luminosity function.

In summary, unlike at \zs4, 3, and 2.2., we are not confident in our
determination of the luminosity function at \zs1.7.  We have chosen to
present the \zs1.7 LF here for completeness, but we urge the reader to
regard it with much caution.

\section{DISCUSSION: THE EVOLVING GALAXY POPULATION}\label{discussion}

\subsection{Summary of the observational evidence}

To summarize \S~\ref{LFevolution}, the number of faint (\subLstar)
galaxies increases with cosmic time, with $\sim$2.3 times more
galaxies of a given luminosity at \zs3 than at \zs4; this increase is
statistically significant at the 11$\sigma$ level and --- as we have
argued in \S~\ref{z4z3faintend} --- it seems very unlikely that it is
an artefact of differential sample selection or cosmic variance in our
data.  At the same time, the evolution from \zs4 to \zs3 appears to be
differential with luminosity, since while the faint end of the LF
evolves significantly the bright end appears to remain virtually
unchanged. This differential effect is statistically significant at
the 97\% level and is unlikely to be due to a systematic bias.  The
case for evolution from \zs3 to lower redshifts is less clear because
of potential systematic biases, although it is possible that the
number density of \subLstar\ galaxies continues to increase to at
least \zs2.2; our analysis suggests that it is extremely unlikely that
the number density of sub-\Lstar\ galaxies at \zs2.2 is {\it lower}
than at \zs3.

Because our results at \zs3 and \zs4 are highly robust (while those at
lower redshifts are less so), in what follows we focus on the
evolution of these two epochs.

\subsection {The importance of luminosity function evolution}

The evolution of the luminosity function measures only the evolution
of the galaxy population {\it as a whole} and does not necessarily
imply a direct correspondence in the evolution of {\it individual}
galaxies. The observed evolution of the LF's faint end can be
interpreted equally well as a change in the number density or
luminosity of the observable population.  However, individual galaxies
are free to change their luminosities following trajectories that are
far more complicated than a direct increase in luminosity or number
density as the evolution of the faint end of the LF might naively
suggest.  Likewise, the apparent constancy of the LF's bright end does
not necessarily imply that the luminous galaxies that populate it do
not themselves evolve.  Evolution of the LF is clearly not a direct
probe of the evolution of individual galaxies.

However, while the luminosity or number density evolution of the LF
does not necessarily reflect a direct corresponding evolution in the
properties of individual galaxies, the fact that the LF {\it does}
evolve constitutes an important suggestion that its constituent individual
galaxies do evolve over time.  Furthermore, the fact that the LF's
evolution appears to be luminosity-dependent suggests that the
evolution of individual galaxies is also differential with luminosity.
It suggests that there may be real physical differences between
low-luminosity and high-luminosity systems in properties such as the
supply of gas available for star formation, the merger rates that may
trigger such star formation, properties of the dust that obscures it,
or the effectiveness of feedback that can regulate it.

\subsection{Some evolutionary speculations}\label{evolutionaryscenarios}

Both semi-analytic and SPH galaxy evolution models have been used to
predict the shape of the LBG luminosity function at different
redshifts (e.g., Somerville, Primack, \& Faber 2001; Nagamine et al.\
2004).  However, such models tend to produce a relatively constant,
unevolving luminosity function --- we are not aware of any predictions
in the literature for the evolution of the \subLstar\ end of the LF
that we observe from \zs4 to \zs3.  It would be interesting to see
what modifications to these sophisticated galaxy formation models can
reproduce a differentially evolving LF.  In the meantime, in the
absence of such predictions, we turn to some simple phenomenological
speculations about the possible nature of the evolution of individual
galaxies that underlies the observed evolution in the LF.  We explore
three heuristic evolutionary scenarios which we use to illustrate how
varying some simple properties of individual galexies can mirror the
observed differential, luminosity-dependent evolution of the LF.

Our three heuristic models are motivated as follows.  SED studies
suggest that high-$z$ galaxies likely undergo episodes of intense star
formation followed by more quiescent periods and it is plausible that
such episodes occur several times in the life of a high-$z$ galaxy
(see, e.g., Sawicki, \& Yee 1998; Shapley et al.\ 2001).  Moreover, it
is clear that UV-selected high-$z$ galaxies are obscured by
significant amounts of starlight-absorbing dust (e.g., Meurer et al.\
1997; Sawicki \& Yee 1998; Ouchi et al.\ 1999; Shapley et al.\ 2001;
Papovich, Dickinson, \& Ferguson 2001; Vijh, Witt, \& Gordon 2003) and
there is no reason to believe that the properties of this dust ---
such as its opacity or large-scale geometry --- remain constant with
time or star formation rate.  Two of our heuristic scenarios
(Scenarios B and C) are thus motivated by the possibility that the
properties of dust or of the starbursting episodes evolve with time
and/or luminosity.  Meanwhile, Scenario A investigates the more simple
picture that in the evolution of the faint end we are seeing the {\it
very first} appearance of many of the low-luminosity galaxies.

We stress that our three heuristic scenarios are not meant as a
comprehensive survey of all possible evolutionary mechanisms.  Clearly
there are many others, but we focus on these three to illustrate some
interesting possibilities and motivate future followup studies.

\subsubsection{Scenario A: The first appearance of low-luminosity galaxies?}

One of the simplest possible pictures of Lyman Break Galaxies is that
they are objects that form their stars at a constant, unvarying rate
for long periods of time.  SED modeling can be used to constrain the
ages of the ongoing episodes of star formation in LBGs and while such
ages are notoriously dependent on the assumed star formation
histories, it is the constant star formation rate scenarios that yield
the oldest ages (Sawicki \& Yee 1998; Papovich et al.\ 2001) thus
providing upper bounds.  Shapley et al.\ (2001) have modeled a large
sample of relatively luminous (typically \R$\sim$24) \zs3 LBGs under
the assumption of constant star formation.  Of 72 luminous \zs3
galaxies in their analysis, only 38\% have ages older than 0.6~Gyr and
so only 38\% of the luminous LBGs seen at \zs3 were forming stars at
\zs4 --- the remainder must have begun their current episodes of star 
formation more recently than \zs4.  Meanwhile, since the number
density of luminous LBGs is the same at \zs4 as at \zs3, 62\% of
luminous \zs4 LBGs must have {\it ceased} star formation before \zs3
to be replaced by the younger starbursts.  Similar reasoning applied
by Iwata et al.\ (2003) to an even earlier epoch suggests that only
20\% of luminous LBGs at \zs5 can still be seen at \zs3 while the
remaining 80\% must have been replaced to keep the bright end of the
luminosity function constant.  These arguments suggest that individual
LBGs cannot be in a steady star-forming state but at best are in a
quasi-steady state where individual galaxies fade in and out of a
given magnitude bin to keep the number density constant, even if
timescales for such fading are long.  Indeed, a detailed analysis by
Ferguson, Dickinson, \& Papovich (2002) calculated that the star
formation histories of (luminous) \zs3 LBGs are inconsistent with the
observed number density of these galaxies at \zs4 unless episodic
bursts of star formation are invoked.

However, the situation may well be different for low luminosity
galaxies --- there it {\it is} possible to reproduce the observed LF
evolution by postulating that a low-luminosity LBG remains at a
constant UV luminosity once it start forming stars.  The age
distribution of \subLstar\ LBGs has not yet been measured, but let us
assume that --- as for the luminous LBGs --- only 38\% of
low-luminosity \zs3 LBGs are old enough to have been present at \zs4.
If at the same time we assume that none of the \zs4 LBGs have ceased
star formation to fade out of the sample, then we can reproduce the
$\sim$2.3-fold number density evolution of low-luminosity LBGs by
assuming that new low-luminosity LBGs are being simply added to the
population between \zs4 and \zs3.

Under this scenario it is possible that we are seeing large numbers of
low-luminosity LBGs ``light-up'' for the first time in the epoch
between \zs4 and \zs3.  If this is the case then we may expect them to
have low metallicities --- a property that should allow us to test
this scenario through future observations.  This picture is perhaps
somewhat akin to ``donwsizing'' scenarios which postulate that
star-formation activity shifts to lower-luminosity objects over time.
It is possible, however, to explain the observed LF evolution with
other evolutionary models as we discuss next.

\subsubsection{Scenario B: Evolution in the properties of starforming bursts?}

The LF evolution may also (or instead) be related to the frequecies,
durations, or intensities of the star-bursting episodes that likely
rule high-$z$ galaxies.  While the constant star formation assumption
in SED fitting of LBGs can result in relatively old starburst ages,
other assumed star formation histories can yield significantly shorter
star-formation episodes (Sawicki \& Yee 1998; Papovich et al 2001) and
it is plausible that such periods of relatively brief elevated star
formation may reccur several times in each galaxy between \zs4 and
\zs3.  Such fluctuating star formation rates are a feature of some 
models of Lyman Break Galaxies (e.g., Nagamine et al.\ 2004).  As the
SFR in a galaxy fluctuates over time, that galaxy will move
back-and-forth between magnitude bins in the luminosity function.  If
the characteristic intensities, durations, or frequencies (duty
cycles) of these episodes decrease with redshift, the resulting effect
will be to alter the shape of the luminosity function.

In particular, if intrinsically low-luminosity galaxies are spending
progressively more time in the state of elevated star formation (be it
because the intense episodes are longer or occur more frequently) then
the faint end of the LF will steepen; an increase in the intensities
(i.e., star formation rates) of the starburst episodes will have a
similar effect on the LF.  At the same time, the bright end of the LF
will remain constant if the characteristics of the starbursting
episodes in the intrinsically luminous galaxies remain fixed with
time.

It is not obvious what mechanism could be responsible for the change
in the duration or intensity of star formation, but one possibility is
that star-formation becomes more robust against self-disruption by
feedback effects as their host dark matter halos accrete material with
time.  The lack of evolution at the bright end would then suggest that
the mechanism depends on halo mass and has saturated for the more
massive galaxies so that it cannot evolve further with time even
though the dark matter halos themselves may still be growing.

\subsubsection{Scenario C: Evolution in the properties of dust?}

The third scenario we examine is linked to the possible evolution in
the properties of interstellar dust in high-$z$ galaxies.  Even a
small change in the amount of dust can have a strong effect on the
observed UV luminosity of a galaxy while leaving its colors relatively
unaffected.  For example a decrease in extinction from \ebv=0.25 to
\ebv=0.1 in a \zs3.5 LBG would produce a six-fold increase in its \Muv\
--- enough to match the evolution of the faint end of the LF we see
between \zs4 and \zs3 --- and yet would result in galaxy colors that
still remain well within the LBG color-color selection criteria.
Because the {\it luminosity} change of the {\it bright} end of the LF
is very strongly ruled out by the data, such evolution in dust
properties would have to be differential with luminosity, which
suggests that such changes cannot be due to differences in sample
selection between redshift bins, but could be due to real changes in
dust properties.

Again, it is not immediately clear what mechanism could result in a
change in effective dust opacity in low-luminosity but not
high-luminosity LBGs.  If the dust evolution scenario is correct than
it may reflect time-dependent changes in the properties of dust grains
in the \subLstar\ LBGs or in the amount of obscured vs.\ unobscured
area visible in each LBG.

\vspace{5mm}

While it is interesting to speculate about the nature of the
underlying evolution of individiual galaxies that is reflected in the
evolving LF, clearly, the luminosity function by itself is
insufficient to discriminate between the possible mechanisms that are
responsible.  To understand what drives the changes we see, we will
have to turn to follow-up studies that compare the properties of dust
and star formation in high-$z$ galaxies as a function of redshift and
luminosity.

\subsection{The way forward}\label{differentialstudies}

A key result is that we have identified luminosity and redshift as
important variables in galaxy evolution at high redshift.  We can use
this fact to seek the nature of the underlying evolutionary mechanism
by comparing diagnostics of dust, age, etc. as a function of $L$ and
$z$.  While LBG follow-up studies to date have primarily focused on
luminous galaxies at z$\sim$3, now that we know that galaxy evolution
depends on $L$ and $z$, extending such studies as a function of
luminosity and redshift provides an attractive way to gain valuable
insights into how galaxies form and evolve.

As we have illustrated in \S~\ref{evolutionaryscenarios}, evolution in
the properties of starbursting episodes or in the amount or
distribution of interstellar dust may underlie the evolution of the
LF.  One line of attack then is to compare the broadband spectral
energy distributions of LBGs as a function of $L$ and $z$.  SED
studies have already yielded insights into the extincton, starburst
ages and stellar masses of LBGs (e.g., Sawicki \& Yee 1998, Papovich
et al.\ 2001; Shapley et al.\ 2001) but by applying these SED-fitting
techniques to different LBG sub-populations we can look for systematic
trends that may reflect the dominant evolutionary mechanisms.
Evolution in the rest-frame UV-through-optical SEDs of {\it luminous
}($L$$\gtrsim$\Lstar) Lyman Break Galaxies from
\zs4 to \zs3 suggests build-up of stellar mass and possibly a 
finely-tuned interplay between increasing dust content and star
formation rates (Papovich et al.\ 2004) even in the observed absence
of evolution in the luminosity function; it will be interesting to see
what SED analyses tell us about the evolution of sub-\Lstar\ LBGs.

Another approach will be to compare the detailed spectra of LBGs as a
function of $z$ and $L$.  A composite spectrum representing a
\R$\sim$24.5 LBG has yielded detailed insights into the properties of
LBG stellar populations, outflows, etc. (Shapley et al.\ 2003).
Comparing composite spectra of LBGs of different luminosity and at
different epochs may yield key insights into what makes LBGs different
as a function of $L$ and $z$.  

Yet another line of attack is to measure galaxy clustering as a
function of both luminosity and redshift as this measurement will let
us relate the potentially time-varying UV luminosity to the more
stable dark matter halo mass; while studies of the luminosity
dependence of clustering have been attempted in the past (Giavalisco
\& Dickinson 2001; Ouchi et al.\ 2004b) they have either relied on vary
small fields or on spectroscopically-untested selection techniques;
the KDF is designed specifically with studying the dependence of
clustering on LBG luminosity in mind and we will attack this issue in
\kdfclustering.

There are two very important advantages that such future differential
studies will have.  Foremost, (1) from an experimenter's viewpoint, we
now {\it know} that $L$ and $z$ are variables that affect how galaxies
evolve.  We can thus be confident that ``varying'' $L$ and $z$ will
yield a ``response'' in galaxy properties linked to evolutionary
mechanisms and that the lack of such ``response'' will equally
importantly rule out a candidate evolutionary mechanism.  At the same
time, (2) while such studies will likely build on previously-developed
techniques as illustrated above, by comparing results as a function of
$L$ and $z$ they will use these techniques in a an essentially {\it
differential} sense thereby reducing our current reliance on
theoretical models or low-$z$ analogs.  Differential measurements are
always much easier and more robust than absolute ones, making such
differential studies extremely attractive.

We feel that important insights lie ahead using this differential
approach and we will pursue such studies in the near future.

\section{SUMMARY AND CONCLUSIONS}\label{summary}

In this paper we have used our very deep KDF catalogs of \UGRI\
color-color selected galaxies at high redhsift to construct the
luminosity functions of UV-selected galaxies at \zs4, 3, 2.2, and 1.7.
As we discuss in detail in \kdfdata, these catalogs use the very same
color-color selection techniques as are used by Steidel et al.\ (1998,
2003, 2004) to select their galaxy samples at these redshifts.
Moreover we use the same effective volume (\Veff) approach to
computing the galaxy luminosity function as used by Steidel et al.\
(1998) at \zs3 and \zs4.  However, our KDF data select galaxies to
\Rlim=27 --- a magnitude and a half deeper than that previous work ---
and so allow us to probe the faint, sub-\Lstar, end of the galaxy
luminosity functions at these and lower redshifts.  Our analysis
probes the population to galaxies that have star formation rates of
$\sim$1\Msun/yr in the absence of interstellar dust and in our assumed
($\Omega_M$, $\Omega_{\Lambda}$, $H_0$) = \cosmoparams\ cosmology.

Spectroscopic redshifts for a large sample of galaxies to
\Rlim$\sim$27 would be observationally extremely expensive, and so
the estimate of the luminosity function at these faint limits {\it
must} at present rely on photometric redshifts or its cousin
color-color selection.  Several attempts to estimate the faint end of
the LF at these redshifts have been made in the past.  However, ours
has the imporant and unique combination of using a color-color
selection technique that is well-understood and well-tested
spectroscopically while at the same time drawing on a galaxy sample
that's taken from a large area of 169\sqam --- thus giving good
statistics --- and spanning three spatially-independent fields ---
thus allowing us to control for cosmic variance.  To our knowledge no
other survey in existence has this combination of favorable and
important characteristics.  We thus believe that ours is the most
reliable estimate of the faint end of the luminosity function at these
redshifts to date.

We have carried out detailed studies to understand the potential
systematic biases in our luminosity function analysis.  We find that
field-to-field variance or uncertainties due to $k$-corrections do not
significantly affect our results and find instead that the largest
source of systematic uncertainty lies in the estimate of the effective
survey volume, \Veff.  We found that our results are robust to how we
estimate \Veff\ at \zs3 and \zs4; however, the estimate of \Veff\
introduces a source of systematic uncertainty into the LF at \zs2.2
and \zs1.7.  Additionally, we also suspect that contamination by
low-$z$ interlopers may be an additional source of uncertainty at
\zs1.7.  Overall, we are highly confident of our LF estimates at \zs4 
and \zs3, we feel we can use the \zs2.2 results to place limits on the
shape of the LF, and are not confident of the \zs1.7 estimate.

In light of the preceding discussion, the results of our analysis can
be summarized as follows:

\begin{enumerate}

\item The faint end slope of the LBG luminosity function at \zs3 and \zs4 
is shallower than the $\alpha$=$-$1.6$\pm$0.13 previously reported at
\zs3 by Steidel et al.\ (1998) using Hubble Deep Field data.  We find
$\alpha$=$-$1.43$^{+0.17}_{-0.09}$ at \zs3 and
$-$1.26$^{+0.40}_{-0.36}$ at \zs4.  While formally consistent with the
Steidel et al.\ (1999) $\alpha$, our more accurate, lower $\alpha$ may
force a factor-of-two downward readjustment of many of the recent
UV-based estimates of the density of star formation in the Universe at
\zs3 and above --- an issue that we address in \kdfsfhist.

\item We find strong evolution in the number density of faint
(sub-\Lstar) LBGs over the 600~Myr from \zs4 to \zs3: there are 2.3
times more sub-\Lstar\ LBGs at \zs3 than at \zs4.  This result is
statistically secure at the 11$\sigma$ level, and we believe it to be
independent of systematic biases due to cosmic variance, sample
selection differences, surface brightness selection differences,
assumptions about the cosmological model, or differential
$k$-correction effects.

\item While the faint end of the luminosity function evolves from \zs4 
to \zs3, the bright end appears to remain unchanged.  This
differential, luminosity-dependent evolution is statistically
significant at the 98.5\% level, where the limitation in our
confidence comes from the small number statistics of the bright end of
the LF.  An improvement in the level of confidence here will require
analysis of \zs4 and \zs3 LBG surveys that are significantly larger
than even the largest that have been studied to date.  If the
differential evolution is real then it may allow new, {\it
differential} approaches to the study of Lyman Break Galaxies.

\item  It is not clear whether the evolution of the faint end of the 
luminosity function continues to lower redshift because of potential
systematic biases at \zs2.2 and \zs1.7.  We find that our estimate of
the \zs2.2 LF depends on our assumptions about the make-up of the
galaxy sample at this redshift.  Despite these systematic
uncertainties, we can nevertheless conclude that there are {\it at
least} as many sub-\Lstar\ galaxies at \zs2.2 as there are at \zs3.

\item At \zs1.7, systematic effects make it diffcult to be confident of 
the reliability of our LF determination at that redshift.

\end{enumerate}

The two most intriguing results of the work presented here are the
increase in the number density of sub-\Lstar\ galaxies from \zs4 to
\zs3 and the possibility of differential, luminosity-dependent evolution 
over that redshift interval.  As we discussed in
\S\S~\ref{systematics} and \ref{z4z3faintend}, the increase in the
number of low luminosity galaxies is a robust result that is both
statistically highly significant and at the same time unlikely to be
an artefact of some systematic bias.  The presence of differential
evolution of the galaxy population is a less secure result (98.5\%
statistical probability, \S~\ref{differentialevolution}) that will
need to be confirmed with much larger, shallow LBG surveys. However,
we note that differential LF evolution is not unexpected given that
there is no reason to think that galaxies across a range of UV
luminosity --- and so, to first order, a range of different star
formation rates --- are straightforwardly scaled analogs of each
other.

The evolution of the faint end of the population raises the intriguing
question: what processes in {\it individual} galaxies underlie the
observed evolution of the population?  A wide range of possible
evolutionary mechanisms may be at play, ranging from changes in the
properties of star-bursting episodes that seem to occur in these
galaxies, to evolution in the amount or properties of interstellar
dust.  Discerning what mechanism is responsible will be important for
our understanding of how high-redshift galaxies form and evolve.  

One avenue of attack on this problem is suggested by the fact that the
evolution of the galaxy population may be differential with
luminosity.  If the population does evolve differentially with
luminosity, then comparing the properties of Lyman break galaxies as a
function of luminosity and (for faint LBGs) of redshift may point us
towards the underlying mechanism.  If --- as some studies suggest
(e.g., Giavalisco \& Dickinson 2001) --- UV luminosity is a tracer of
the dark matter halo mass, then we will first be able to link the
differential evolution of the population to a mass scale.  We are
starting to pursue this line of attack with our KDF data
(\kdfclustering).  Another set of insights will be possible from
comparing the spectral energy distributions of LBGs or the details of
their composite spectra; such studies have to date been focussed on
relatively luminous LBGs at \zs3, where they have yielded insights
into, e.g., the starbust ages, extinction, and the state of their
interstellar media (Sawicki \& Yee 1998; Papovich et al.\ 2001;
Shapley et al.\ 2001; 2003).  The intriguing possibility of
luminosity-dependent evolution of the LBG population opens up the
possibility for such studies in a way that is {\it differential} and
so largely independent of the systematics associated with using models
or low-redshift analogs.  We will explore such approaches in upcoming
work.


\vspace{5mm}

We thank the Palomar time allocation committee for a generous time
allocation that made this project possible and the staff of the W.M.\
Keck Observatory for their help in obtaining these data.  We are
grateful to Chuck Steidel for his encouragement and support of this
project and Jerzy Sawicki for many useful comments.  We also thank
Masami Ouchi and Armin Gabasch for providing their LF data points in
tabular format and Naveen Reddy and Ikuru Iwata for useful
discussions.  Finally, we wish to recognize and acknowledge the very
significant cultural role and reverence that the summit of Mauna Kea
has always had within the indigenous Hawaiian community; we are most
fortunate to have the opportunity to conduct observations from this
mountain.

\newpage



\clearpage




\clearpage

\clearpage

\clearpage

\clearpage

\clearpage

\clearpage

\clearpage

\clearpage

\clearpage

\clearpage

\clearpage

\clearpage


\begin{deluxetable}{lcccc}
\tablewidth{0pt} 
\tablecaption{\label{schechter.tab}Parameters of Schechter Function Fits}
\tablehead{
\colhead{z} & 
\colhead{Steidel type\tablenotemark{a}} & 
\colhead{$M^*_{1700}$} & 
\colhead{log $\phi^*$\tablenotemark{b}} & 
\colhead{$\alpha$}
}
\startdata
1.7 & BM & 		 $  -19.80^{+0.32}_{-0.26} $     & $-1.77^{+0.11}_{-0.11}$    & $-0.81^{+0.21}_{-0.15}$ \\ 
2.2 & BX & 		 $  -20.60^{+0.38}_{-0.44} $     & $-2.52^{+0.20}_{-0.26}$    & $-1.20^{+0.24}_{-0.22}$ \\ 
3   & C, D, M, and MD &  $  -20.90^{+0.22}_{-0.14} $     & $-2.77^{+0.13}_{-0.09}$    & $-1.43^{+0.17}_{-0.09}$ \\ 
4   &  & 		 $  -21.00^{+0.40}_{-0.46} $     & $-3.07^{+0.21}_{-0.33}$    & $-1.26^{+0.40}_{-0.36}$ \\ 
\enddata
\tablenotetext{a}{In the nomenclature of Steidel et al.\ 2003, 2004}
\tablenotetext{b}{In units of Mpc$^{-3}$}
\end{deluxetable}

\begin{deluxetable}{lcccccc}
\tablewidth{0pt} 
\tablecaption{\label{phistar_by_field.tab}Number density $log$ \phistar\tablenotemark{a} in different fields of the survey}
\tablehead{
\colhead{z} & 
\colhead{combined fields\tablenotemark{b}} &
\colhead{02A}&
\colhead{03A}&
\colhead{03B}&
\colhead{09A}&
\colhead{09B}
}
\startdata
1.7 & $-1.77^{+0.11}_{-0.11}$ & $-1.92$ &  $-1.80$ &  $-1.80$ &  $-1.70$ &  $-1.74$    \\ 
2.2 & $-2.52^{+0.20}_{-0.26}$ & $-2.59$ &  $-2.41$ &  $-2.44$ &  $-2.65$ &  $-2.55$    \\ 
3   & $-2.77^{+0.13}_{-0.09}$ & $-2.79$ &  $-2.69$ &  $-2.88$ &  $-2.77$ &  $-2.74$    \\ 
4   & $-3.07^{+0.21}_{-0.33}$ & $-2.97$ &  $-3.24$ &  $-3.27$ &  $-3.16$ &  $-2.94$    \\ 
\enddata
\tablenotetext{a}{In units of Mpc$^{-3}$}
\tablenotetext{b}{From fit to the full data as reported in Table~\ref{schechter.tab}}
\end{deluxetable}

\end{document}